\input lecproc.cmm
\contribution{Dissipation in Relativistic Outflows:\newline
A Multisource Overview}
\author{Christopher Thompson}
\address{Physics and Astronomy, University of North Carolina, \newline
Chapel Hill, NC 27599}
\abstract{Relativistically expanding sources of X-rays and $\gamma$-rays
      cover an enormous range of (central) compactness and Lorentz factor.
      The underlying physics is 
      discussed, with an emphasis on how the dominant dissipative mode
      and the emergent spectrum depend on these parameters.  
      Photons advected outward from high optical depth are a potentially
      important source of Compton seeds.  Their characteristic energy
      is bounded below by $\sim 1$ MeV in pair-loaded outflows of
      relatively low compactness, and remains near $\sim 1$ MeV at
      very high compactness and low matter loading.  This is compared
      with the characteristic energy of $O(1)$ MeV observed in the
      rest frame spectra of many sources, including $\gamma$-ray bursts,
      OSSE jet sources, MeV Blazars, and the intense initial 0.1 s pulse
      of the March 5 event.  Additional topics discussed include the
      feedback of pair creation on  electron heating and the formation of
      non-thermal spectra, their effectiveness at shielding the dissipative
      zone from ambient photons, direct Compton damping of irregularities
      in the outflow, the relative importance of various soft photon sources,
      and the softening of the emergent spectrum that results from heavy
      matter loading.  The implications of this  work for X-ray and optical
      afterglow from GRB's are briefly considered.   Direct synchrotron
      emission behind the forward shock is inhibited by the
      extremely low energy density of the ambient
      magnetic field.   Mildly relativistic ejecta off axis from the
      main $\gamma$-ray emitting cone become optically thin to scattering
      on a timescale of $\sim 1$ day $(E/10^{52}~{\rm erg})^{1/2}$, and
      can be a direct source of afterglow radiation.}

\titlea{1}{Introduction:  Variety of Sources and Spectral Behavior}

X-ray and $\gamma$-ray emission from relativistic outflows is
powered by the conversion of bulk kinetic energy and Poynting luminosity,
by a variety of possible mechanisms.  However, the assumed values of
key parameters such as the Lorentz factor $\gamma$
of the outflow, the compactness of the central source, as well
as the optical depth and size of the
dissipative zone, vary dramatically between the different classes of sources.
For example, $\ell_c \sim 10^{15}$ and $\gamma\sim 10^2-10^3$ 
are inferred for cosmological
$\gamma$-ray burst (GRB) sources vs. $\ell_c \sim 10-10^2$ and $\gamma 
\sim 3-30$ for Blazars (Fig. 1).
This motivates a more global analysis of how dissipation of kinetic and
magnetic energy is achieved, which provides some interesting new perspectives
on particular sources.   

\begfig 7 cm
\includegraphics{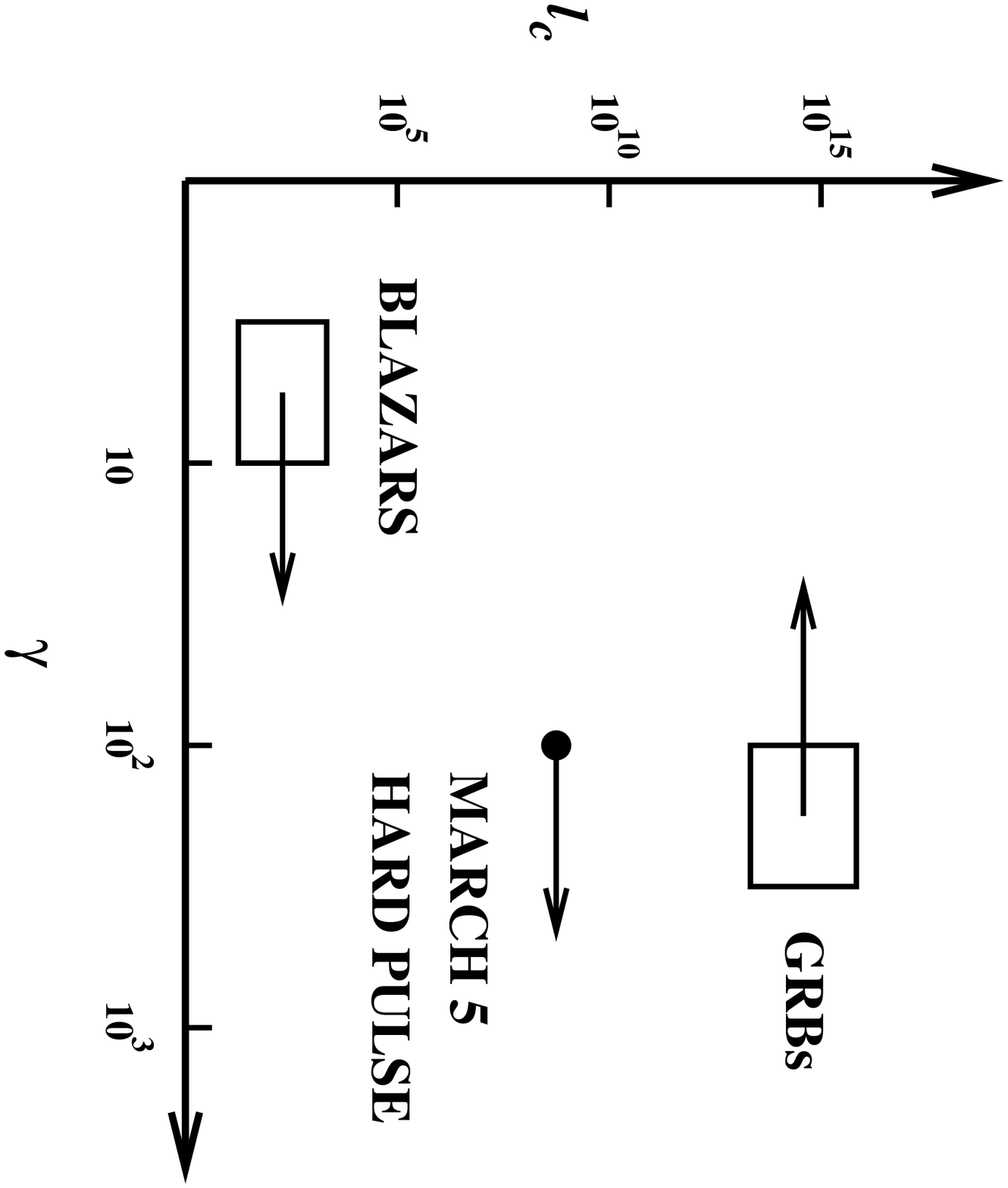}
\figure{1}{Blazars, $\gamma$-ray burst sources and
the initial 0.1 sec hard pulse of the March 5, 1979 burst occupy
distinct regions of the plane defined by source compactness $\ell_c$
and asymptotic Lorentz factor $\gamma_\infty$.  GRB outflows may contain
lower-$\gamma_\infty$ ejecta that manifests itself as softer
tails, precursors and sub-pulses (T94, T96).  
The Lorentz factors of Blazar sources
are only indirectly constrained by superluminal motion outside the
$\gamma$-ray emitting region (cf. Wagner, these proceedings).}
\endfig

Another key point is that sources (or types of sources) exhibit
different {\it spectral states}.  GRBs are predominantly
non-thermal, but some 
contain subluminous precursors, tails, and sub-pulses with 
distinctly thermal high energy cutoffs (Yoshida et al. 1989;
Pendleton et al. 1996).  Blazars are occasionally observed
with emission peaked strongly at $\sim 1$ MeV, in distinction to the
more usual extended power-law behavior 
(Bloemen et al. 1995; Blom et al. 1995). And the remarkable 5 March 1979
burst was initiated by an extremely intense $\sim 0.1$ sec flare whose
luminosity exceeded that of the remainder of the burst by a factor
$\sim 300$ and showed much more pronounced spectral softening
(Fenimore et al. 1996).  

Spectral states with sharp high energy cutoffs have a simple interpretation
as the residue of an {\it optically thick} outflow.  Indeed, at very high
$\ell_c$, the outflow is self-shielding from the central radiation source,
as well as from external (e.g. side-scattered) radiation.   
Radiation advected outward from large scattering depth then becomes an
important source of seeds for extended non-thermal spectra -- 
in addition to the more familiar optically thin synchrotron-self-Compton
mechanism.  This means that the inner boundary conditions on the flow
are much more important than is usually supposed.   And this raises
an interesting question:  are sources of relatively low compactness
(e.g. Blazars) ever self-shielding in this manner?

It is intriguing to note, in this regard, that the characteristic energy
$\sim 1$ MeV appears in a number of sources (of widely varying parameters):
the spectral breaks observed in OSSE jet sources ($\sim 1-3$ MeV after
compensating for cosmological redshift; e.g. McNaron-Brown et al. 1995)
and in classical GRBs ($\sim 100$ keV $-1$ MeV at peak luminosity, 
without compensating for redshift; Mallozzi et al. 1995);  in MeV Blazars;
as well as the initial hard spike of the March 5 burst ($\sim 300$ keV).
Of course, the possibility that selection effects narrow the
observed spectral break distribution should be considered 
carefully.  This happens in the case of the GRB sources 
(Piran \& Narayan 1996) only if the total burst energy is constrained to
yield an inverse correlation between break energy and flux -- in distinction
to the strong positive correction observed within individual bursts.
Most plausible cosmological GRB sources release as much as $10^{53}-10^{54}$
erg, which allows for a fraction of very energetic and hard bursts.

After careful consideration of various photon sources, it turns out that
an advected Wien peak maintains a characteristic energy of $\sim 1$ MeV 
over a wide range of central compactness, if the flow is sufficiently
relativistic (Sects. 1.2; 3.1).   Pair creation
by photon collisions, $\gamma + \gamma \rightarrow e^+ + e^-$,
which has traditionally been viewed as inimical to high energy gamma-ray
production, can in fact play an essential role by i) increasing the
efficiency of leptonic dissipative modes; ii) reducing the lower cut-off energy
of the non-thermal pair distribution to $\gamma_{min} \sim 1$, 
which yields a break in the spectral distribution of Compton-upscattered
Wien photons near the position of the original Wien peak; and iii)
selecting non-thermal over thermal spectra at Comptonizing hotspots
a high-$\gamma$ outflow (Sect. 3.3).  
Since the photon collision cross section is
comparable to Thomson, feedback from pair creation works most effectively
near $\tau_T \sim 1$.  This contrasts with the inhomogeneous
external Comptonization model (Blandford and Levinson 1995, hereafter
BL95), where the non-thermal high energy continuum emerges well outside
the scattering photosphere.  Indeed, the position of the scattering
photosphere in a pair-loaded outflow is sensitive to the amount of
continuous heating, and pair creation can significantly broaden
the transition zone between optically thick and thin flows (Sects. 1.2, 2.2).

Thus, by focussing on those aspects of the physics that are special to the
large-$\ell_c$ GRB regime, and then considering how these vary with
compactness, interesting new insights can be obtained on both the
GRB and Blazar problems.  When constructing GRB models, it is sobering to
realize that Blazars are still far from being understood, even with the
much broader spectral information available.

In these notes, I will explore the following additional points:

$\bullet$  The dependence of the dominant dissipative mode on optical depth.
 At $\tau_T > 1$ direct
Comptonization by bulk fluid motions is most effective; whereas 
non-thermal (e.g. Fermi) particle acceleration is a crucial
ingredient of any radiative model at low $\tau_T$.  
Strong-wave acceleration can be excluded if enough scattering charges
are present to generate an observable flux of Comptonized
high energy photons.

$\bullet$  What is the relative importance of double Compton emission
and cyclo-
synchrotron emission as seeds for Comptonization (at large $\ell_c$)?

$\bullet$ How does the influence of geometrical effects (e.g. beaming)
vary with $\gamma_\infty$?

$\bullet$ High energy cutoffs to extended power-law spectra are
extremely diagnostic.  In GRB sources these are very poorly constrained.
Measurements by EGRET in the 30 MeV - 10 GeV range indicate
a significant decorrelation with the 1 MeV flux, with the high energy
emission often being significantly {\it delayed} (Hurley et al. 1994).

\titleb{1.2}{The $\ell_c$-$\gamma_\infty$ Plane}

The variety of possible dissipative modes is neatly summarized in
a two-dimensional plane labeled by (Fig. 2)
$$\ell_c = {L_{rel}\sigma_T\over 4\pi m_ec^3 R_0};\;\;\;\;\;\;\gamma_\infty = 
{L_{rel}\over \dot Mc^2}.\eqno(1)$$
The radius $R_0$ of the central engine is identified with the
Alfv\'en radius or light-cylinder radius as appropriate.
At risk of oversimplification I will usually assume that
$\gamma$ has attained the limiting value $\gamma_\infty = 
L_{rel}/\dot Mc^2$ due to matter loading $\dot M$ (when discussing
delayed dissipation at large distances from the central engine).

\begfig 7 cm
\includegraphics{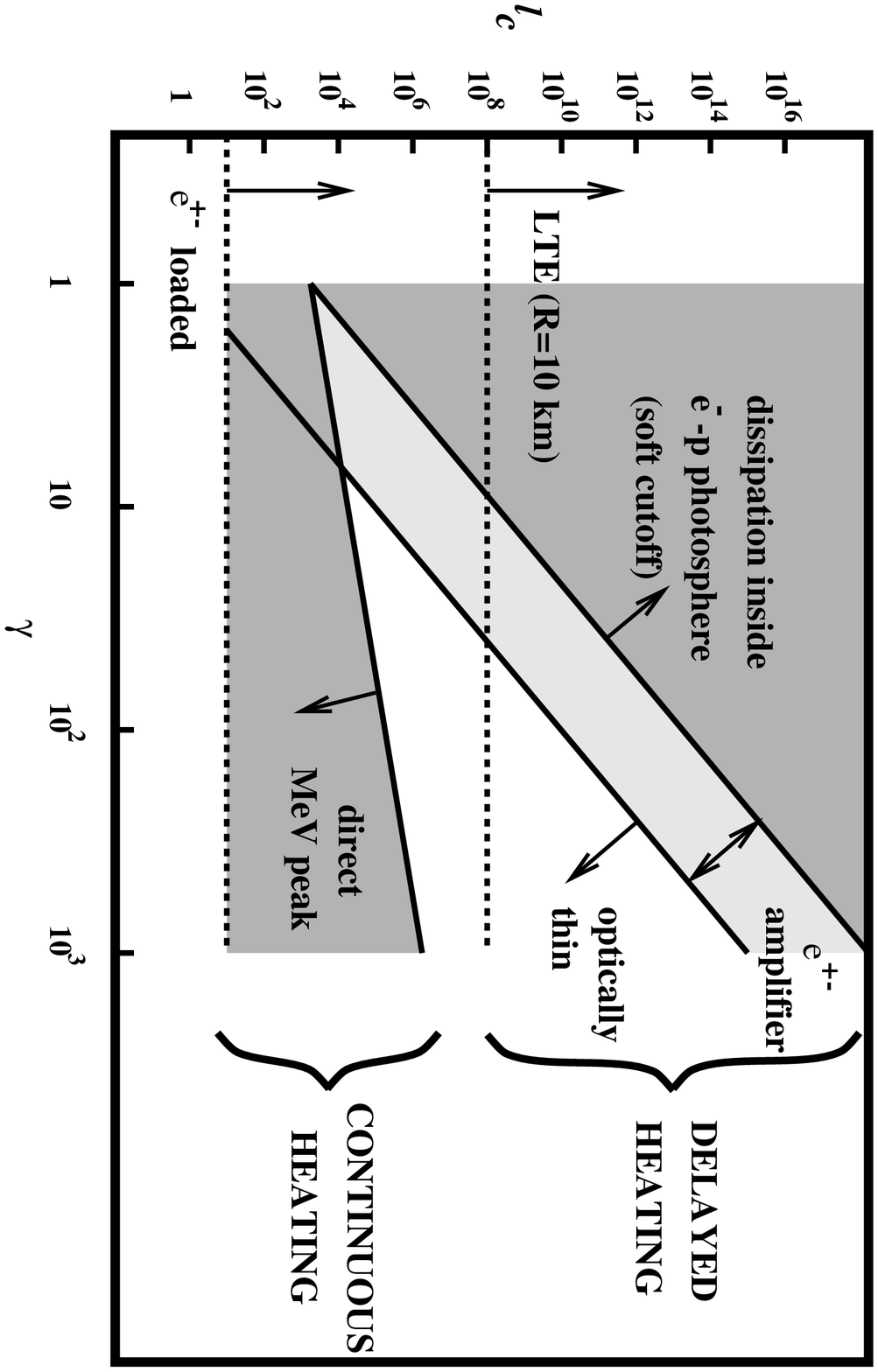}
\figure{2}{The variety of dissipative regimes in relativistic outflows
as summarized in the $\ell_c$-$\gamma_\infty$ plane.  Flows
with heavy matter loading and asymptotic Lorentz factor 
$\gamma_\infty < \gamma_{e-p}$ occupy the top left region.  They are
a plausible source of soft subcomponents of GRBs.
Flows with $\gamma_{e-p} < \gamma_\infty <  \gamma_{e^\pm}$ can become
pair loaded, and occupy the adjoining stripe.  Flows with low
matter loading, $\gamma_\infty > \gamma_{e^\pm}$, dissipate at low
optical depth.   Above the upper horizontal dashed line, 
local thermodynamic equilibrium is achieved at the base of the flow;  
whereas below this line the photon distribution is Wien.
All of the flows in this region are assumed to 
undergo {\it delayed} dissipation after thermal pairs freeze out,
at a radius where inhomogeneities (such as reconnecting magnetic
fields and shocks) regain causal contact.   By contrast, flows in the
lower portion of the diagram are assumed to dissipate {\it continuously}
and remain pair loaded out to the scattering photosphere.   This is
expected in collimated flows with lower asymptotic Lorentz factors (such
as Blazars).  Large scattering depths cannot be maintained below the lower
horizontal dashed line.}
\endfig

The radius at which dissipation takes place is limited significantly
by causality at large $\gamma$, as long as 
$\gamma(r)$ grows faster with radius than $r^{1/2}$
near the base of the outflow.  Thus, it is convenient to define the
compactness $\ell_{\Delta t} = \ell_c\times (R_0/c\Delta t)$ associated with
variability on a timescale $\Delta t$.   A plausible value of the dissipative
radius is $\sim 2\gamma_\infty^2 c\Delta t$
for a variety of dissipative modes, including magnetic reconnection and
MHD turbulence (Romanova \& Lovelace 1992, hereafter RL92; 
Thompson 1994, 1996, hereafter T94, T96; Levinson, these proceedings) 
and shocks powered by variations in $\dot M$ (Pacyz\'nski \& Xu 1994, hereafter
PX94; Rees and M\'esz\'aros 1994, hereafter RM94).
The radius of the scattering photosphere of the wind is related to $R_{diss}$
by a very strong function of $\gamma_\infty$,
$$
{R^{e-p}_{\tau=1}\over 2\gamma_\infty^2 c\Delta t} \sim 
{m_e\over m_p}{\ell_{\Delta t}\over \gamma_\infty^5}\;\;\;\;\;\;(n_p \gg 
n_{e^+}),\eqno(2)
$$
when the scattering depth is dominated by the advected electron-ion
contaminant.  A non-thermal photon tail extending above energy $m_ec^2$
in the rest frame will greatly increase the density of scatterers, with
the result
$$
{R^{e^\pm}_{\tau=1}\over 2\gamma_\infty^2 c\Delta t} \sim 
{\ell_{\Delta t}\over \gamma_\infty^5}\;\;\;\;\;\;(n_{e^+} \gg n_p)\eqno(3)
$$
for a photon index $\beta \sim -2$ characteristic of GRBs.  The effects
of pairs are discussed further in Sect. 3.

Outflows with Lorentz factor $\gamma_\infty < \gamma_{e-p} = 
(m_e/m_p)^{1/5}\ell_{\Delta t}^{1/5}$ dissipate well inside the
electron-ion photosphere.  They occupy the upper left portion of
Fig. 1, and have a possible association with the soft X-ray precursors,
tails and quasi-thermal sub-pulses of GRBs (T96).    
Outflows with $\gamma_{e-p} < \gamma_\infty 
< \gamma_{e^\pm} = \ell_{\Delta t}^{1/5}$
dissipate inside the the pair photosphere, if the high energy continuum
extends up to $\sim m_ec^2$ in the rest frame.
And, outflows with $\gamma_\infty > \gamma_{e^\pm}$
dissipate at low scattering depth, independent of the efficiency of
pair creation.  
The spectral consequences of variations in the matter
loading are discussed further in Sect. 3.

Flows in which the irregularities maintain causal contact 
will undergo continuous heating while optically thick.  An interesting 
example of such a flow is the low-$\gamma$ sheath of a 
relativistic jet, in which the luminosity of entrained photons increases
with radius as the kinetic energy of the higher-$\gamma$ core of the jet
is dissipated.  Although 
the relation between scattering depth and internal temperature
of the flow is sensitive to the high energy distributions of the pairs
and photons,  the mean energy of the emergent photon spectrum is regulated to
near $\sim\gamma m_ec^2$.  In the case where the photon distribution is Wien, 
the strong $T$-dependence 
$n_{e^\pm}/n_\gamma = (\pi/2)^{1/2}(T/m_ec^2)^{-3/2}\exp(-m_ec^2/T)$
of the equilibrium pair density  guarantees that as soon as $T$ drops
much below $m_ec^2$ in the rest frame, the flow becomes optically thin.
This yields a simple relation between the observed (Lorentz-boosted) 
temperature and the temperature $T_0$ at the base of the flow,
$$T_{obs} = T_0\,(L_{\gamma}/L_{\gamma\,0}),\eqno(4)$$
in terms of the growth of the photon luminosity from $L_{\gamma\,0}$
to $L_\gamma$.  

The pair
density resulting from direct heating of the photons by MHD turbulence 
can be less sensitive to $T$.  I assume that the bulk of the
photon energy lies in a Wien bump, but that wave energy is
excited in transient surges (e.g. by reconnection) with an equivalent
temperature $T_w$ somewhat in excess of $m_ec^2$.  Then a significant
fraction of the wave energy is converted to photons above the pair
production threshold.  Balancing this source against pair annihilation,
the scattering depth in the direction perpendicular to a jet (of opening
angle $\theta$) is $\tau_{T\perp} = {1\over 2}n_e\sigma R\theta \sim 
(\gamma\ell_\gamma)^{1/2}$, where
$\ell_\gamma$ is the compactness (1) in the advected photons at radius $R$.
At $\tau_T > 1$ ($\ell_\gamma > 1$), 
freshly created pairs Compton cool before annihilating (the cooling
timescale being shorter by a factor $\sim \tau_{T\parallel}^{-1}$)
and carry a fraction $\sim \tau_{T\parallel}^{-1}$ of the total energy of
the flow.  
This regulates the Compton parameter induced by mildly relativistic pairs
to $y \simeq \tau_{T\parallel}^{-1}\cdot\tau_{T\parallel} \sim 1$.  The
annihilation photons Compton downscatter off the cooled pairs
to an energy $\sim m_ec^2/\tau_{T\parallel}$ in the rest frame. 
The photon spectrum emerging at the pair photosphere then peaks
at an energy $\sim \gamma m_ec^2$.
This implies only a modest increase in the mean energy per photon
along the jet, by a factor $\sim \gamma m_ec^2/3T_0$,
which is easily supplied from the high-$\gamma$ core to the low-$\gamma$
sheath.

Continuous heating by relativistic electrons at low optical depth has
been considered by Sikora et al. (1997) as a model for the MeV Blazars.  
They relate the peak energy to the observed variability timescales, but
since this energy is not directly tied to a microphysical scale, it could
be expected to lie well below $\sim 1$ MeV in some sources.
By contrast, direct Compton damping of mildly relativistic turbulence
in an optically thick jet yields a Wien peak energy that is bounded below
by $\sim 1$ MeV (although still dependent on bulk $\gamma$ and the
energy transferred from the bulk motion to the photons).

\titlea{2}{Dissipative Mechanisms}

\titleb{2.1}{Sources of Free Energy}

The internal sources of free energy in a relativistic flow can be broadly 
divided into two categories:  those associated with radial and angular
inhomogeneities.  Both appear to be important in jet sources, and
while both have also been considered in GRB sources, radial inhomogeneities are
probably more important in the large-$\gamma$ context.  Reconnection surfaces
and variations in the ratio of particle pressure to magnetic pressure 
will lead to internal heating of strongly-magnetized outflows
(RL92; T94), as will kinetic energy
fluctuations in particle-dominated outflows (PX94; RM94).
However, interactions with an external medium (Rees and 
M\'esz\'aros 1992) can become significantly non-spherical as the outflow
in a $\gamma$-ray burst source decelerates (Sect. 4).

Which of these energy sources dominates depends on the strength of the
magnetic field in the outflow and the radius of the dissipative zone.
It appears that $\gamma_\infty \sim 100-300$
can be achieved in an outflow of luminosity $\sim 10^{51}$ erg s$^{-1}$
only if it is Poynting-flux dominated at the
source.  Indeed, {\it any} triggering scenario for a GRB that produces
an object with the density of nuclear matter and a rotation period of
$\sim 10^{-3}$ s plausibly involves magnetic fields as strong as
$\sim 10^{15}$ G through dynamo amplification and thus leads to 
a rotationally-driven
MHD outflow of luminosity $L_P \sim 10^{50}-10^{51}$ erg s$^{-1}$ (T94;
see also Duncan \& Thompson 1992; 
Usov 1992, 1994; Vietri 1996; M\'esz\'aros and Rees 1997
for particular models).  The alternative mechanism of
$\nu-\bar\nu$ annihilation into $e^\pm$ pairs has an efficiency of
$\sim 10^{-3}$ (Jaroszynski 1993), and so the neutrino luminosity
required to power a $\gamma$-ray flux of $\sim 10^{51}$ erg s$^{-1}$
drives a mass flux (by absorption on nucleons) that exceeds the tolerable
value by $\sim 10^6$ (cf. Duncan, Shapiro and Wasserman 1986).  

This advected magnetic field almost certainly has an important
effect on the dissipative mechanism.
Quasi-perpendicular shocks are significantly weakened even when
the magnetic field carries a few percent of the energy flux;  this
in turn steepends non-thermal particle spectra arising from
first-order Fermi acceleration.   Although particle acceleration at
oblique relativistic shocks can be quite efficient (as discussed by
Kirk, these proceedings), one expects that {\it internal} shocks in
GRB outflows with $\gamma \sim 100-300$ are approximately radial.
Thus, GRB models based on internal shocks may unfortunately require
complicated departures from spherical symmetry.  

\titleb{2.2}{Geometrical Effects:  Pair Cocoons}

The huge compactness of a GRB source ($\sim 10^{15}$) causes
it to be {\it self-shielding}.  Not only is the central engine
hidden from the dissipative zone, but the optical depth
$\tau_\perp$ perpendicular to the axis of the flow exceeds the parallel
depth by the very large factor,
$${\tau_\perp\over\tau_\parallel} \sim \gamma^2\theta \sim 10^{4-5}\theta,
\eqno(5)$$
for reasonable values of the opening angle $\theta$.

This raises the question:  as $\ell_c$ decreases toward the values 
typical of AGN, at what point does the central engine become visible?
One intriguing possibility is that the engine remains shielded in some Blazars,
with the high-$\gamma$ cores of the jet being surrounded by a 
{\it pair cocoon} (Fig 3). Because Lorentz dilation causes $\tau_\parallel$ to
decrease in proportion to $\gamma^{-2}(L_{jet}/\dot M_{jet} c^2)^{-1} 
\propto \gamma^{-3}$, the
high-$\gamma$ core of a jet can become optically thin along its axis
well inside the photosphere of the $\gamma \sim 1-2$ sheath.  This
lies at a radius
$$R_{\tau= 1}^{e^\pm} \sim 2\times 10^{18}\,\left({L_{jet}\over 10^{47}~
{\rm erg~s^{-1}}}\right)\,\left({\theta\over 0.1}\right)^{-1}\;\;\;\;\;\;
{\rm cm}, \eqno(6)$$
assuming that the sheath acquires a non-negligible fraction of the kinetic
energy at this radius.    All that
is required (Sect. 1) is that the sheath be i) pair-loaded and optically
thick at its base, and ii) {\it continuously} heated (e.g. by Kelvin-Helmholtz
instabilities with higher-$\gamma$ material) at a sufficient rate
to overcome the effects of adiabatic cooling.  Condition ii) is
plausibly satisfied in the cores of superluminal sources, and
i) is satisfied if the outflow is strongly turbulent at its base (Sect. 3).  
Below this threshold heating rate,  the position of the pair photosphere
is very sensitive to the amount of heating.

\begfig 7 cm
\includegraphics{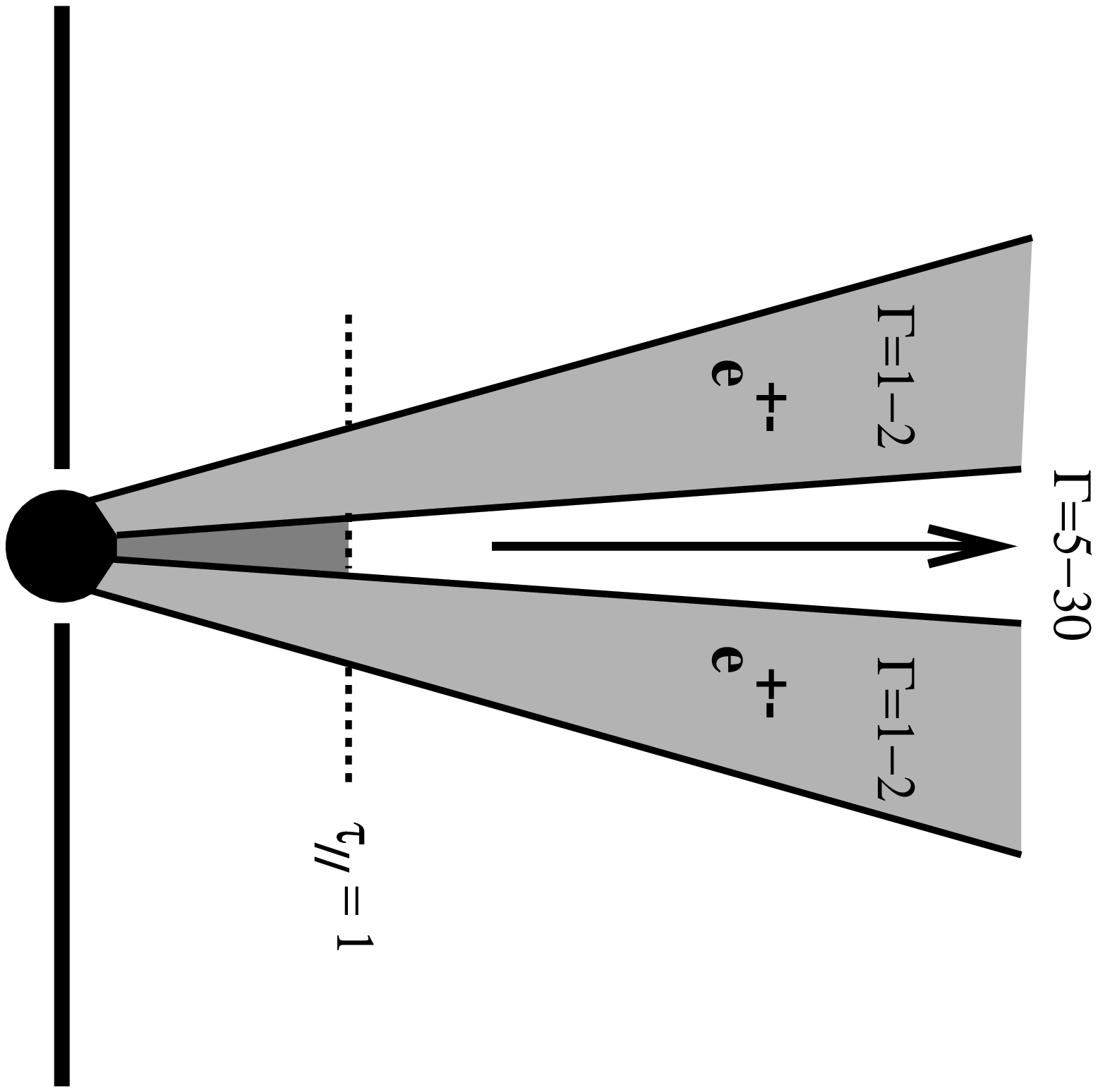}
\figure{3}{A sheath of low-$\gamma$ material surrounding the
high-$\gamma$ core of a jet can remain pair loaded out to a large
distance (6) from the central engine.  The scattering photosphere
of this {\it pair cocoon} will in general lie far outside the radius
at which the core becomes transparent to high energy $\gamma$-rays.}
\endfig

\titleb{2.3}{Diverse Photon Sources}

When the flow is self-shielding in this manner, {\it advected} 
quasi-thermal radiation
becomes an important source of Compton seeds (T94, T96).
Even though external radiation exerts a stronger drag force
than internally generated radiation by a factor $\gamma^2$
(Sikora, Begelman, \& Rees 1994; BL95), it
cannot penetrate the high-$\gamma$ component of the flow.
This also disfavors models involving acceleration
and heating of the jet by central continuum radiation propagating
along the jet axis (Dermer and Schlieckeiser 1993; 
Marcowith et al. 1995).
The main competing source of seed photons is then synchrotron
radiation (e.g. Maraschi et al. 1992; M\'esz\'aros, Rees and 
Papathanassiou 1994, hereafter MRP94).

In quantifying the relative importance of these two radiation
sources, one must consider separately thermal and non-thermal scatterers.
The advected radiation characteristically has a much higher frequency
than the cyclotron frequency, and if its energy density $U_\gamma^a$
is comparable to $B^2/8\pi$, then it will dominate the
magnetic field as a coolant of thermal electrons, due to self-absorption
near the cyclotron resonance.   Parametrizing the frequency at
which the cyclo-synchrotron radiation becomes self-absorbed in terms
of a cyclotron harmonic $N_{sa}$, one deduces from Kirchoff's law that
C-S cooling rate is smaller than the Compton cooling
rate off the advected radiation by the factor
$${\nu j_\nu \over 4\sigma_T n_e c(T/m_ec^2)
U_\gamma} = {2\alpha_{em}\over\pi} {N_{sa}^3\over \tau_T}\,
\left({B\over B_{QED}}\right)\,{B^2/8\pi\over U_\gamma},\;\;\;\;\;\;
\left(\nu = N_{sa} {eB\over 2\pi m_e c}\right)\eqno(7)
$$
where $\alpha_{em} = 1/137$ and $B_{QED} = 4.4\times 10^{13}$ G.  For
example, one expects $B/B_{QED} < 10^{-8}$  
near the scattering photospheres of GRB outflows, so that
C-S cooling can be neglected even if $N_{sa} \sim 300$. 

Another effect of the advected radiation is to limit the Compton
parameter $dy/d\ln R \simeq 4(T/m_ec^2)(1+4T/m_ec^2)d\tau_T/d\ln R$ in 
optically thick regions of the flow.  The magnitude of this effect
depends on the mean energy per photon $\langle h\nu\rangle$, 
and the bulk Lorentz factor $\gamma$.  If the flow is photon rich,
with $\langle h\nu\rangle/\gamma \ll m_ec^2$ in its rest frame,
then the advected photons limit $y$ to a value $\sim \ln(\langle \nu\rangle/
\nu_0)$, where $\nu_0$ is peak frequency before the advected photons
undergo (delayed) reheating.  This prevents very low energy C-S photons
from being upscattered much in frequency.  A stronger limit
$y \sim 1$ applies in outflows that are continuously Compton heated by
MHD waves and shocks (Sect. 1.2).   Nonetheless,
a much larger $y$-parameter will be maintained at the
base of the outflow, where the photon flux has a net divergence (Sect. 3.1).

Synchrotron radiation from Blazar sources is usually ascribed to
the same high energy electrons/positrons that are responsible for the
X-ray/$\gamma$-ray emission.  The high energy cutoff of the 
synchrotron peak has been ascribed to a suppression of the non-thermal
particle density inside the scattering photosphere (e.g. Levinson 1996),
but mildly relativistic pairs are only suppressed by a factor
$\sim \tau_T^{-1}$.  Thus, the particle spectrum must itself
steepen considerably at $\tau_T > 1$.
Rather large minimum non-thermal Lorentz factors $\gamma_{min}$ are sometimes
conjectured in GRB sources, sufficient to place the synchrotron
peak energy directly in the MeV range (e.g. MRP94).
However, synchrotron absorption becomes much
more important in that context, if the outflow
becomes sufficiently pair loaded that $\gamma_{min} \sim 1$ (Sect. 3.3).

Advected radiation can also be the dominant coolant at {\it external}
relativistic shocks.  Consider a shell of relativistic matter of
initial width $c\Delta t$.  After the shell becomes optically thin
to scattering, the radiation moves ahead of the shell, but continues
to overlap inside a radius $R > 2\gamma^2\,c\Delta t$.  The point is
that when the external medium is cold (with a sound
or Alfv\'en speed much less than $c$), the external magnetic field $B_{ex}$
that is swept up and compressed in between the forward shock and
the contact discontinuity typically has a {\it much} smaller energy
density than the radiation, by a factor
$${B^2/8\pi\over U_\gamma} \sim 2\times 10^{-10}\,
\left({B_{ex}\over 3\times 10^{-6}{\rm ~G}}\right)^2\,
\left({L_\gamma\over 10^{50}~{\rm erg~s^{-1}}}\right)^{-1}\,
\left({\gamma\over 10^2}\right)^8\,\left({\Delta t\over 10~{\rm s}}\right)^2
\eqno(8)
$$
at $R = 2\gamma^2c\Delta t$ (assuming a compression factor of 7).
{\it This strongly suggests that direct synchrotron radiation from 
the forward shock is {\rm not} the mechanism primarily
responsible for delayed optical and X-ray emission from GRBs.}
Direct radiation from a low-$\gamma$ component of the ejecta is
considered in Sect. 4.

\titleb{2.4}{Dissipation at $\tau_T > 1$}

MHD waves, turbulent motions and shocks can in principle tap a
significant fraction of the energy of the outflow.  A periodic excitation of
frequency $\omega_k$ involving a displacement $\xi_k$ of the fluid
deposits its energy {\it directly} in the photons via Compton drag 
when the scattering depth across $\xi_k$ is less than $\sim c/\omega_k\xi$
(T94).  For example, relativistic Alfv\'en waves have a damping time 
$$t_{drag} \omega_k \sim {(\delta B)^2_k/8\pi\over U_\gamma},\eqno(9)$$
which is shorter than the wave period as long as the photon gas has
a higher energy density than waves {\it in the relevant range of wavenumbers}
(Thompson and Blaes 1997, hereafter TB97).
This mechanism is particularly effective near the scattering photosphere,
and is also effective even at large $\tau_T$,
as long as higher wavenumber turbulence is generated via a turbulent
cascade.
The wave amplitude generally decreases with wavenumber in such a cascade,
with the result that $(\delta B)_k^2/8\pi \ll U_\gamma$ at dissipative
scales even if $(\delta B)_k^2/8\pi \sim U_\gamma$ at the outer scale.
Shocks also transfer energy to the photon fluid via compression and direct
first-order Fermi acceleration (Blandford and Payne 1981), 
although energy transfer is slowed significantly when the photon pressure
becomes comparable to the material ram pressure ahead of the shock.

\titleb{2.5}{Dissipation at $\tau_T < 1$}

The high energy emission in Blazars covers a wide range of energies,
up to 10 GeV (or higher in the TeV sources) and hence
must be powered by non-thermal particle distributions.   The most familiar
possibilities are first-order Fermi acceleration at shocks (Blandford
and Eichler 1987) and electrostatic acceleration (e.g. RL92).
Photo-pion production on protons can more easily 
accomodate the TeV sources (Mannheim 1993), but requires a supplementary
$\sim$ MeV emission mechanism in sources 
with soft high energy spectral states, because of the much greater
cross-section for $\gamma + \gamma \rightarrow e^\pm$.

The relevant physics is treated in sufficient depth elsewhere that
I will focus on two questions here.

1. {\it Does reconnection deposit energy primarily in thermal or
non-thermal particles, and in electrons or ions?}  This problem is
far from being understood from first principles, but Solar flares
do provide clear evidence that the efficiency of electron acceleration
can (at least in a non-relativistic plasma) be quite high.  However,
Type III radio bursts (which are powered by flare particles that
escape the Sun along open magnetic field lines) also provide direct evidence
that most of the flare energy is dissipated in the form of {\it bulk
heating} of electrons to energies of 10-100 keV, with only a small
fraction being deposited in a relativistic, non-thermal tail (Lin 1990).

2. {\it What is the effectiveness of electrostatic and
strong-wave acceleration in a relativistic, Comptonizing medium, 
as compared to shock acceleration?}  This question highlights
a crucial difference between the large scale relativistic outflows
associated with Blazars and GRBs, and a laboratory system such as a
Tokomak (or even smaller scale astrophysical systems such as
coronal loops and arcades).  A magnetic field $B$ with gradient scale
$\ell_B = B/|{\bf\nabla} B|$ requires a minimal charge density
$n_{c,min} = B/4\pi e\ell_B$ to support the associated current; otherwise
the displacement current cannot be neglected and charges are accelerated
to relativistic energies.  The ratio of the actual electron density
to $n_{c,min}$ can be expressed in terms of the scattering depth across
$\ell_B$,
$${n_e\over n_{c,min}} \sim {\tau_T\over\alpha_{em}}\,
\left({B\over B_{QED}}\right)^{-1}.\eqno(10)$$
This works out to $n_e/n_{c,min} \sim 10^{13}$ for parameters appropriate to
inhomogeneous external Compton Blazar models (BL95).
It has been hypothesized that the boundary layers of jets may
be partially evacuated and sites for strong-wave acceleration 
(Bisnovaty-Kogan \& Lovelace 1997), but in fact the degree of evacuation
must be extraordinary for such a mechanism to be important.  

Another approach to this problem
is to re-express $B$ in terms of the plasma $\beta_e=8\pi n_eT/B^2$,
$${n_e\over n_{c,min}} \sim \beta_e\left({B\over B_{QED}}\right)\,
\left({T\over m_ec^2}\right)^{-1}\,{\ell_B\over \ell_{m_e}},\eqno(11)
$$
where $\ell_{m_e} = 4\times 10^{-11}$ cm is the Compton wavelength
of the electron.  This is $n_e/n_{c,min} \sim 10^4\beta_e$,
$10^8\beta_e$, and $10^{21}\beta_e$ for parameters appropriate to 
Tokomaks, Solar flares, and Blazar jets.  This suggests that 
much higher wavenumber distortions of the magnetic field are
required to provide efficient electron acceleration through
reconnection in relativistic outflows.
By contrast, direct Comptonization of a background photon
fluid is {\it more} effective in jets and GRBs due to the
higher scattering depth (T94).

\titlea{3}{Spectral Consequences}

Most modelling of high energy emission from relativistic outflows 
ignores the inner boundary condition on the outflow.
The flow is hypothesized to dissipate outside the pair
annihilation radius, and flow conditions interior to that radius are
assumed not to influence the emergent high energy spectrum.  Seed
radiation for Comptonization is assumed to originate in a central
accretion disk {\it exterior} to the volume of the outflow.

We have already seen, however, that the emergent spectrum can 
be dominated by advected radiation if the outflow is optically
thick at its base (T94, T96).  Moreover, the inner boundary conditions are
very well defined in flows of large central compactness, such
as GRBs ($\ell_c \sim 10^{15}$).   The mean photon energy emerging from 
the flow is $\langle h\nu\rangle \sim L_P/\dot N_\gamma$, when the
asymptotic Lorentz factor lies near the critical value $\gamma_{e^\pm}$
(or $\gamma_{e-p}$ if pairs are absent). Thermalization is rapid near
the base of the flow,  the photon gas is very close to black body, and 
$\langle h\nu\rangle$ is directly related to the effective temperature
at the light cylinder, 
$$
\langle h\nu\rangle \sim T_{eff} = 
0.8\,\left({L_{\gamma}\over 10^{50}}\right)^{1/4}\,
\left({P\over{\rm 10^{-3}~s}}\right)^{-1/2}\;\;\;\;\;\;{\rm MeV}.\eqno(12)
$$
This is remarkably close to the observed range of spectral break
energies, after allowing for cosmological redshift.  

In this regime, the ratio of photon luminosity $L_\gamma$ 
to (ordered) Poynting luminosity $L_P$ at the base of the wind is a key
 parameter.  It should be emphasized that the 
baryon loading is tolerably small only if $L_\gamma < 10^{-2} L_P$ at
the neutrinosphere.   In other words, a key requirement of this model is that 
the wind be reheated from $L_\gamma \ll L_P$ to $L_\gamma \sim L_P$
well outside the neutrinosphere.  This is plausibly accomplished by
a MHD cascade to high wavenumber, even though photons and pairs
are tightly coupled on macroscopic scales (Sect. 2.4; TB97).

Expression (12) also leads to an interesting question:  how does the mean
energy per photon change as one decreases the central compactness?  
As we now show, the trend
of decreasing mean photon energy with decreasing compactness in fact
can be reversed, with $\langle h\nu\rangle$ approaching $\sim m_ec^2$
at $\ell_c \sim 10^2$.

\titleb{3.1}{Direct MeV Wien Peak}

A magnetized outflow that is strongly turbulent will trigger a
cascade to high wavenumber that {\it must} dissipate inside the Alfv\'en
radius.  We look for an optically thick equilibrium state in which the
dissipative zone is shielded from external photons and soft photons are
generated internally.  In applications to AGN the advected matter contributes
negligible optical depth, and so the outflow is necessarily hot and
pair loaded.   Damping via resonant couplings between high wavenumber
turbulence and cosmic ray particles has been considered by Dermer, Miller,
\& Li (1996).  However, direct Compton drag of bulk turbulent motions
is an effective damping mechanism at high compactness, typically at much
lower wavenumbers (TB97).  I now estimate the temperature $T_0$
of the flow at its base, focussing on two photon sources.

{\it 1. Double Compton Emission.}  This dominates bremsstrahlung emission
when $T_0 \sim m_ec^2$ and the outflow is photon rich, $n_\gamma \gg n_e = 
n_{e^+} + n_{e^-}$.  In a Wien photon gas,
$$\dot n_{dC} = {16\Lambda\over\pi}\alpha_{em}n_en_\gamma\sigma_T c
\left({T_0\over m_ec^2}\right)^2,\eqno(13)
$$
where $\Lambda \simeq \ln(T/h\nu_{min})$ and the photon gas approaches
a Planckian distribution at frequency $\nu_{min}$.

Double Compton dominates cyclotron emission when the magnetic
field in the central engine exceeds $B_{QED} = 4.4\times 10^{13}$ G, 
so that thermal pairs do not populate excited Landau levels.
Such strong fields have indeed been associated with SGR 0526-66, which emitted
the March 5, 1979 superburst.  The initial 0.1 s pulse of that burst
appears to have approached a luminosity of $\sim 10^7$ times the Eddington
luminosity (Fenimore et al. 1996), and has the appearance of an expanding
pair fireball (Thompson \& Duncan 1995; Fatuzzo \& Melia 1996).  
If the outflow is driven by a magnetic field that is also strongly
turbulent, then Compton drag can raise the photon energy density 
$3T n_\gamma$ close to $B^2/8\pi$ near the light cylinder (TB97).
Equating $\dot n_{dC}$ with the photon loss rate,
one deduces an equilibrium scattering depth
$$\tau_T = {\pi\over 16\Lambda\alpha_{em}}\left({T\over m_ec^2}\right)^{-2}.
\eqno(14)$$
Re-expressing this in terms of the equilibrium pair density yields
the relation between $T_0$ and compactness $\ell_c$ shown in Fig. 4.

\begfig 7 cm
\includegraphics{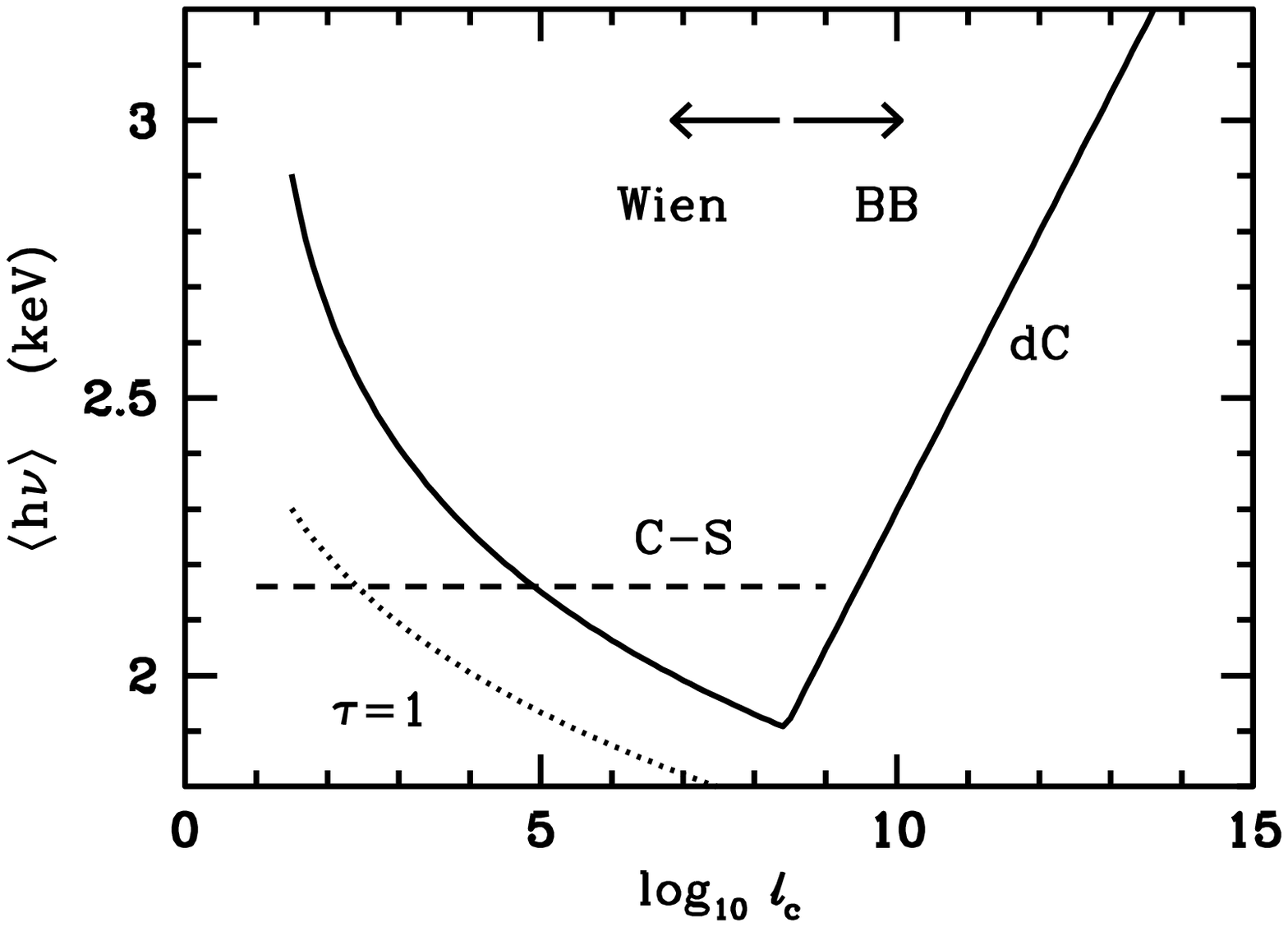}
\figure{4}{Mean photon energy $\langle h\nu\rangle$ versus central
compactness $\ell_c$
for flows in which double Compton emission (solid line) and
cyclo-synchrotron emission (long-dashed line) is the dominant soft photon
source.  The minimum temperature for an optical thick flow (assuming
a Wien photon distribution) is labelled by the short-dashed line.  Note
the breaks in the curves at the transition from Wien to black-body photon
distributions.}
\endfig

{\it 2.  Cyclo-Synchrotron Emission in a Non-thermal Pair Plasma}.
Such a very high optical depth and compactness cannot be maintained
in weaker magnetic fields.  Cyclo-synchrotron photons are created rapidly,
and their energy rapidly exponentiates.  (For related calculations
with non-thermal particle distributions, see Ghisellini, Guilbert, \&
Svensson 1988.)  To show this, I
parametrize by $N_c eB/m_ec$ the critical frequency at which Compton
scattering increases the frequency of a C-S photon at the same rate as it is
absorbed, $dy/dt = c\alpha_\nu$.  From Kirchoff's law,
$${\dot n_{C-S}\over \dot n_{dC}} \simeq {2N_c^2\over\Lambda}\,\left({B^2/8\pi
\over n_\gamma m_ec^2}\right).\eqno(15)$$
Since $N_c \gg 1$ for $3T \sim m_ec^2$  (Mahadevan, Narayan, \& Yi 1996)
this yields $\dot n_{C-S} \gg \dot n_{dC}$.  

Is the energy released by a turbulent cascade
at the base of the outflow dissipated at large or small optical depth?
If even a tiny fraction of the bulk kinetic energy is carried by particles,
then the wave energy must cascade to very high wavenumber before 
charges are accelerated either electrostatically -- as the turbulence
becomes charge starved -- or by resonant interactions.  The relative 
importance of these two effects is determined by the ratio
$${k_z\over eB/m_ec^2} \sim \left({\ell\over\ell_{m_e}}\right)^{-1}\,
{\tau_T\over\alpha_{em}}\,\left({B\over B_{QED}}\right)^{-2}$$
in the case of sheared Alfv\'en waves of wavenumber $k_z$ (TB97).
The cascade can, however, be cut off by Compton drag much closer to
the outer scale.  The MHD wave motions are only mildly relativistic,
and so a large scattering depth  is required to provide the Compton parameter
$y \sim 10$ needed to upscatter the bulk of the C-S photons to high energy.
The mean energy per photon is given by
$${3T_0\over m_ec^2} \simeq {B^2/8\pi\over m_ec^2 \dot n_{C-S}(R/c)} \sim 
{\pi\over 32 N_c^2\alpha_{em}} \tau_T^{-1}\,
\left({T_0\over m_ec^2}\right)^{-2},\eqno(16)$$
where I approximate the bulk of the photon distribution as Wien.   
Assuming that
thermal and turbulent velocities are comparable, one has
$T_0/m_ec^2 \simeq y/8\tau_T$ and
$$\tau_T \sim 13\,\left({y\over 10}\right)^{3/2}\,\left({N_c\over 20}\right).
\eqno(17)$$
This gives $T_0/m_ec^2 = 0.095\,(y/10)^{-1/2}(N_c/20)^{-1}$, as shown
in Fig. 4. 

This scattering depth lies above the thermal value, and so must be
maintained by an extended non-thermal tail to the pair distribution 
function.  This is naturally provided by MHD wave heating, as discussed
in Sect. 1.2.  The minimum central compactness
needed to support this optically thick state is, on energetic grounds,
 $\ell_c \sim 10$ (Fig 2).


\titleb{3.2}{Dissipation at Large Compactness and Low $\gamma_\infty$}

A much wider range of peak energies is possible if enough matter 
is advected by the outflow to make it optically thick.   This is possible
only for a rather higher compactness, $\ell_c > m_p/m_e \sim 2000$,
than is expected in Blazar sources, but well within the range of
cosmological GRB sources (cf. Paczy\'nski 1990).  
If the matter loading is heavy enough that $\gamma_\infty 
< \gamma_{e-p} = 0.2 \ell_{\Delta t}^{1/5}$, then dissipation
on an observed timescale $\Delta t$ occurs inside the $e-p$
scattering photosphere.   This provides a natural explanation for the
soft tails and precursors to GRBs seen by Ginga, and the soft sub-pulses
seen by BATSE (T94, T96).

Unlike pair-dominated outflows with negligible matter (Sect. 3.1)
the emergent temperature is very sensitive to the amount of continuous
heating.  (In the case of spherical flows with radial inhomogeneities,
this is equivalent to a broad power spectrum of inhomogeneities.)
I further assume that the flow expands relativistically at its base,
with $\gamma(r) \propto r^\alpha$ with $\alpha > {1\over 2}$.   Then
inhomogeneities in the flow fall out of causal contact,\fonote{In
contrast to the model of MeV Blazars discussed in Sects. 2.2 and 3.1,
in which the collimated, lower-$\gamma$ 
flow is assumed to be heated continuously 
by non-radial Kelvin-Helmholtz instabilities.}  and dissipation
on timescale $\Delta t$ is {\it delayed} to radius $R_{dis} \simeq 
2\gamma^2c\Delta t$.  The corresponding luminosity $L(R_{dis})$
is then depleted by adiabatic expansion out to the scattering photosphere.
Assuming that photon-number changing processes freeze out before dissipation
takes place, the emergent temperature is a very strong function of matter
 loading,
$${T_{obs}\over T_0} = 0.7\,{L(R_{dis})\over L_\gamma(R_0)}\,
\left({\gamma\over \gamma_{e-p}}\right)^{10/3}\;\;\propto\;\;
(\Delta t)^{2/3}.\eqno(18)$$
The correction for adiabatic losses (the last factor in this expression)
has a very strong dependence on $\gamma$, but a much weaker dependence
on the (more directly observable) variability timescale $\Delta t$.
This means that the power spectrum of inhomogeneities must cover a very 
wide range of frequencies for continuous heating to overcome the effects of
adiabatic cooling in a {\it radial} flow.  The situation is quite
different in collimated flows, where energy can be continuously
extracted from angular (e.g. Kelvin-Helmholtz) instabilities (Sect. 2.2).

The number of advected photons must also be compared with
the number of fresh cyclo-synchrotron photons generated in the
flow outside the central engine.  Following Sect 3.1, this is
$${\dot n_{C-S} t\over n_\gamma} \sim {8\alpha_{em} N_{sa}^2\over \pi \gamma}
{d y\over d\ln t}\,\left({T\over m_ec^2}\right).\eqno(19)
$$
Here all quantities refer to the rest frame of the outflow.
The advected photons suppress $\dot n_{C-S}$ at large $\tau_T$
by soaking up energy from the electrons and holding down $y$.  We
conclude that $\dot n_{C-S}/n_\gamma$
is typically less than unity for high-$\gamma$ outflows.

\titleb{3.3}{$e^+-e^-$ Amplifier in Relativistic Outflows}

The efficiency with which the available energy (in shocks, MHD waves, and
reconnecting magnetic fields) is deposited in high energy photons depends
directly on the fraction $\varepsilon_e$ of the energy deposited in
electrons and pairs.  Pair creation (via photon collisions $\gamma + \gamma
\rightarrow e^+ + e^-$) has traditionally been used to constrain
the emission region in high energy sources 
(e.g. Cavallo \& Rees 1978; Baring \& Harding 1997), but one would like
to emphasize an opposing point of view here:  that the high energy photon flux
from a relativistic outflow can be significantly {\it raised} by
pair creation, due to an increase in $\varepsilon_e$.   Energization
of positrons by high-harmonic proton synchrotron maser radiation behind a shock
(Hoshino \& Arons 1991) provides an example of such a leptonic
acceleration mechanism.

Pair creation has, of course, been included for some time in models
of BH accretion disk coronae (e.g. Stern et al. 1996 and
references therein), but in that context the formation of a power law
high energy continuum does not depend essentially on the presence of pairs.
For example, direct Comptonization by MHD motions in the corona will
effectively heat the photons, independently of the 
relative amounts of rest energy in baryons and leptons (T94).  
The situation is quite different in a relativistic outflow that expands
sufficiently rapidly at its base that inhomogeneities fall out of causal
contact.  If these inhomogeneities cover a wide range of spatial frequencies
$k_{min} < k < k_{max}$ (as is needed in GRB models to accomodate the
broad power spectra of the bursts) then pair creation at wave number
$k_{max}$ (radius $\sim 4\pi\gamma^2k_{max}^{-1}$) will increase the
radius of the scattering photosphere by a factor 
$${R_{\tau=1}^{e^\pm}\over R_{\tau=1}^{e-p}} \sim
\left({m_p \over m_e}\right)\,
\left({E_{br}\over\gamma m_ec^2}\right)^{2-\beta},\eqno(20)$$
given a high energy photon index $\beta$.  This {\it pair amplifier}
allows a much wider range of wavenumbers to be dissipated directly by
Compton drag, and hence causes regions of the wind with non-thermal
high energy spectra to be significantly brighter than regions with
thermal spectra (T96). 

At this point I should distinguish between pair amplification 
that is {\it linear} (the probability of energization $P_e$ of a newly
created pair is the same as that of a seed electron) and {\it non-linear}
(the pairs feed back on $P_e$).  For example, the amplifier operating
at a shock is non-linear if fresh pairs all act as suprathermal 
seeds
for first-order Fermi acceleration; whereas it is linear if the pairs
cool down to the temperature of the background thermal plasma before
interacting resonantly with plasma waves.\fonote{The gyroperiod of a
relativistic electron is orders of magnitude shorter than its cooling
time if the magnetic field contributes an appreciable fraction of the
pressure of the outflow.}   The
corresponding leptonic efficiencies are
$$\eqalign{
\varepsilon_e &= {P_e + 2n_{e^+}/n_p\over P_p+P_e + 2n_{e^+}/n_p}\;\;\;\;\;\;
({\rm non-linear});\cr
&= {P_e(1 + 2n_{e^+}/n_p)\over P_p + P_e(1+n_{e^+}/n_p)}\;\;\;\;\;\; 
({\rm linear}),\cr
}\eqno(21)$$
where $P_p$ is the probability of energization of a proton and
$\varepsilon_e = P_e/(P_e +P_p)$ in the absence of pairs. 

The pair amplifier operating at Comptonizing hotspots (Sect. 3.4)
is non-linear in different sense:  pair creation regulates
the high energy index $\beta$ to the appropriate value to yield
a Thomson depth $\tau_T \sim {1\over 4}(T/m_ec^2)^{-1}$ 
within individual hotspots.  
This second-order Fermi acceleration mechanism therefore yields a power-law
high energy spectrum over a wider range of matter loadings
($\gamma_{e-p} < \gamma_\infty < \gamma_{e^\pm} = 4.5\gamma_{e-p}$) than
does synchrotron cooling of shock-accelerated pairs.

Pair amplification via photon collisions is also {\it non-local}:  
photons upscattered above energy $m_ec^2$ in one hotspot will raise
the pair density in another portion of the flow.  A nice example 
is provided by an expanding shell of matter and photons of radial width
$c\Delta t$.  The bulk Lorentz factor of the photon gas is not
limited by the inertia of the matter when $\gamma > \gamma_{e^\pm}$.
Outside a radius $\sim 2\gamma^2c\Delta t$ 
photons with energies greater than $m_ec^2$ in
the wind frame stream ahead of the forward shock, and sidescatter against
seed electrons.  The pair density grows exponentially (at first)
inside a radius $\sim \ell_c R_0$, until $n_{e^+}/n_p$ reaches unity
and the material ahead of the shock is accelerated to a limiting
Lorentz factor $\sim \ell_\gamma^{1/2}$ (for a hard incident photon
spectrum with $\beta = -2$).  The feedback of pair creation on the
structure of a relativistic shock is an interesting problem that has
not been properly addressed.

One immediate spectral consequence of the pair amplifier is a 
suppression of the minimum leptonic Lorentz factor $\gamma_{min}$,
and hence a suppression of the minimum synchrotron frequency\fonote{The
feedback of pair creation on the formation MeV breaks in high energy
synchro-Compton cascades above accretion disks has been considered
by Done, Ghisellini and Fabian (1990).}
$E_{sync}({\rm min}) \sim \gamma_{min}^2 eB/m_ec$.  
In fact, $\gamma_{min} \rightarrow 1$
as the inertia of the pairs becomes comparable to that of the protons.
To give an example of the potential importance of this effect, consider
a Poynting-flux dominated outflow that approaches its limiting
Lorentz factor $\gamma_\infty$.  The high energy photon index is
taken to be $\beta = -2$ out to a rest frame energy $m_ec^2$.  Near the
scattering photosphere of the wind, the cyclotron energy is
$$\hbar{eB\over m_ec} = {8\pi ec^2\over \sigma_T (2L_Pc)^{1/2}}
\,\gamma_\infty^3 =  0.04\,\left({\gamma_\infty\over 300}\right)^3\,
\left({L_P\over 10^{51}~{\rm erg~s^{-1}}}\right)^{-1/2}\;\;\;\;{\rm eV}.
\eqno(22)
$$
The efficiency of electron acceleration can be increased by pair
creation, but at the cost of suppressing $E_{sync}({\rm min})$ far below
the observed range of break energies in GRB spectra.  {\it As a result, 
the primary emission process must be inverse Compton.}

\titleb{3.4}{Delayed Inhomogeneous Comptonization:  \newline
Broken Power-law Spectra with a Thermal Photon Source}

Let us now consider the photon spectrum that results from delayed
reheating of a relativistic outflow at large $\ell_c$, 
outside the electron-ion photosphere ($\gamma_\infty \sim \gamma_{e-p}$).
At large $\ell_c$, the photons are adiabatically cooled in between the
central engine and the causal contact radius,
where they are reheated to a luminosity $L_\gamma \sim (\delta B/B)^2L_P$
(when the outflow is Poynting-flux dominated). 
The mean photon energy is restored to a value 
$$\langle h\nu\rangle \sim 0.7 L_\gamma/L_{\gamma 0},\eqno(23)$$
given that photon number is conserved at this radius (Sect. 3.2).

As before, we consider a broad power spectrum of inhomogeneities,
$k_{min} < k < k_{max}$; the corresponding (radial) size of a hot spot
is $\Delta \sim \pi/k$.  Let us suppose that wavenumbers $k_\star < k
< k_{max}$ dissipate inside the electron-ion photosphere, and
$k_{min} < k < k_\star$ outside.  Individual hotspots are assumed to
release their energy when the causal propagation distance
$R/2\gamma_\infty^2$ begins to exceed $\epsilon^{-1}\cdot\Delta$.

Hotspots with $k > k_\star$ dissipate when the
scattering depth of the flow is $\tau_T  = k/k_\star$
(due to seed electrons). The scattering depth across an individual spot
$\tau_T^{spot}$ is smaller by $\varepsilon$.  The resultant spectrum
is Wien when $\tau_T^{spot} \gg 1$.  When $\tau_T^{spot} < 1$
(but the flow itself is still optically thick) cold seed photons
escape the spot before being upscattered, and one may use the
standard loss-probability formalism (Shapiro, Lightman, \& Eardley 1976).
Since the seed photons have adiabatically cooled by a factor
$\sim (2\gamma_\infty)^{-2/3}(kR_0/2\pi)^{2/3}$, the accumulated $y$-parameter
required to upscatter them is large, and the resulting photon index is
$$\alpha = {1\over 2} - \sqrt{(9/4) + (4/y)} \simeq -1.\eqno(24)$$
This power law distribution [extending up to a mean energy (23)],
with a superimposed Wien peak at energy (23), is the net
result of this first stage of Comptonization.  It compares favorably
with the low energy spectra of GRBs (e.g. Cohen et al. 1996).

As dissipation continues at wavenumber $k \sim k_\star$,
Compton drag regulates the $y$-parameter to a value near unity.
Hotspots with temperature $T_w \sim m_ec^2$ will 
upscatter photons above energy $\langle h\nu\rangle$ in a non-thermal tail
that extends to the pair creation threshold in the wind rest frame.  
This in turn greatly amplifies the number of scattering charges,
since photons greatly outnumber electrons in the outflow (by
a factor $\sim \gamma_\infty (m_p/m_e) (m_ec^2/\langle h\nu\rangle)$.  

The key point here is that the resulting expansion of the
scattering photosphere feeds back directly on the shape of the high
energy continuum (T96).  If the photon 
compactness $\ell_\gamma = L_\gamma\sigma_T/4\pi\gamma^3 m_ec^3 R \gg 1$
at this radius, then a high energy photon index as hard as $\beta = -2$
generates $\tau_T \gg 1$ within individual hotspots, which
in turn prevents the formation of an extended high energy continuum.
For example, if the heating is triggered by reconnection (T94) then
this requires only that $V_A \sim c$ in the wind rest frame,
so that individual reconnection events induce bulk mass motions at
velocities close to the speed of light.  As the compactness drops,
$\beta$ rises to maintain $\tau_T \sim {1\over 4} y (T_w/m_ec^2)$ 
within individual hotspots.\fonote{
As long as $\epsilon \ll 1$ photons are able to diffuse freely between
spots, but not escape the wind entirely.}  In other words, the feedback
works primarily through the scattering depth, rather than through
a balance between the time-averaged heating and cooling rates ($y = 1$)
as in accretion disk corona models (Shapiro et al. 1976; 
Haardt \& Maraschi 1993).

The net result is that hotspots in the wind with 
the right properties to generate pairs are observed to be much
brighter in X-rays and $\gamma$-rays than are
regions of the wind
with thermal spectra, because a much wider range of wavenumbers
is dissipated by Compton drag (Sect. 3.3).

This mechanism does not require fine-tuning of the Lorentz factor if 
the range of wavenumbers is broad, $k_{max} \gg k_{min}$.  Nonetheless,
one expects that $\gamma_\infty$ is a strong function of time
in any GRB source involving an optically thick neutron torus or
neutron star that emits neutrinos (T94).  The  neutrino luminosity
plausibly passes through the critical value at which the
neutrino driven mass-loss rate is $\dot M = L_P/c^2\gamma_{e-p}$;
indeed the total Poynting luminosity $L_P$ from a centrifugally
supported torus is limited to $L_P \sim 10^{51}$ erg s$^{-1}$ in
this manner.  

A related model uses strong shocks to directly accelerate
the photons via the first-order Fermi process (Blandford and Payne 1981).
If the photons pass successively through several strong shocks
separated by adiabatic cooling, then it can be shown that the number
index converges to a value $-1$ (Melrose \& Pope 1993) up to an energy
$\sim L_P/\dot N_\gamma$.

\titleb{3.5}{A Hybrid Model:\newline
Comptonization of an MeV Bump by Non-thermal Pairs}

An advected Wien photon gas with temperature $T_0 \sim m_ec^2$ can
seed Comptonization by non-thermal pairs below the scattering
photosphere.   If the distribution of relativistic pairs has a lower cutoff
$\gamma_{min} = O(1)$, then {\it the resultant spectrum
breaks in the MeV range}.  Such a low cutoff results from
a high energy pair cascade in a compact photon source (BL95), and 
results even for steeper pair spectra if pair creation feeds back
on the leptonic acceleration efficiency to load the outflow heavily
with pairs, $n_e m_ec^2 \sim (\delta B)^2/8\pi$ (Sect. 3.3).  

A further benefit of heavy pair loading is that the outflow becomes
photon-starved when $3T$ approaches $m_ec^2$ in the rest frame,
so that the advected bump is strongly depleted by Compton upscattering above
$\sim 1$ MeV.   The resultant break energy is then (for $\nu F_\nu \sim $
const above 1 MeV)
$$h\nu_{br} \sim {L_\gamma\over \dot N_\gamma}\,
\left[\ln\left({h\nu_{max}\over m_ec^2}\right)\right]^{-1}
\sim 3T_0\,\left({L_\gamma\over L_{\gamma 0}}\right)\,
\left[\ln\left({h\nu_{max}\over m_ec^2}\right)\right]^{-1}.\eqno(25)
$$
Indeed, a narrow bump near 1 MeV appears to be the exception rather
than the rule in Blazar spectra, although PKS 0208-512 does provide
a spectacular exception (von Montigny et al., these proceedings).

As a model for Blazar spectra, this has a number of advantages over
models involving i) direct Comptonization of photons from the central
source (Dermer and Schliekieser 1993); and ii) Comptonization of
side-scattered photons (Sikora, Begelman, \& Rees 1994; BL95).
First, a Comptonized UV bump (Sikora et al. 1997) is avoided because
the flow is self-shielding
(Sect. 2.2); second, the advected Wien peak has a high enough temperature that
the photon source is depleted during creation of the power-law
$\gamma$-ray spectrum; and, third, observations of {\it both} MeV 
power-law breaks and isolated MeV bumps in Blazar spectra are
directly tied to the electron rest energy.  {\it The duality between these
two spectral states is ascribed to the presence or absence of a strong
non-thermal $e^\pm$ component.}

Comptonization of an advected MeV bump can be powered either by thermal or
non-thermal particles.  Is it reasonable to expect that quasi-thermal
motions should be the dominant Compton heat source in GRB sources,
but non-thermal particles in Blazars?   The key difference between these
sources, aside from the central compactness, appears to be 
the degree of relativistic
expansion.  Shocks in a GRB outflow should (locally) more closely approximate
spherical surfaces, with the result that first-order Fermi
acceleration is strongly suppressed in the relativistic limit (cf.
Kirk, these proceedings).

\titlea{4}{Conclusions:  Optically Thick vs. Thin Sources}

We have studied dissipation in relativistic outflows that
are sufficiently compact ($\ell_c > 10$) to be optically
thick at the center.  Advected MeV radiation  provides an
interesting new source of Compton seeds in this regime.  Interacting with
bulk turbulent motions and non-thermal pairs {\it near the scattering
photosphere}, it can manifest itself either as an MeV Blazar, or as an extended
power law state when most of the available energy is converted to
non-thermal pairs.  The absence of a prominent MeV bump
in most Blazar sources can be explained since the outflow is
{\it photon starved}.   The spectral signature is expected to be
different when $\ell_c < 1-10$, or when the advected radiative flux is low.
The high energy spectral break energy can cover a wider range of
frequencies when synchrotron photons are the dominant seeds
(MRP94; Ghisellini, these proceedings; Takahara, these proceedings).

Electron-positron pairs play a crucial role here by
i) maintaining a large scattering depth near the base of the outflow
and shielding the high-$\gamma$ core of a jet from ambient radiation;
ii) maintaining the mean energy of the advected radiation above $\sim 1$ MeV; 
iii) enhancing the efficiency of leptonic dissipative modes;  and iv)
reducing the minimum energy of the non-thermal pair population to 
$\gamma_{min} \sim 1$, which keeps the minimum energy of the Comptonized
MeV photons in the MeV range.  In a large-$\gamma$ (GRB) outflow,
pairs also feed back on the emergent spectrum
by expanding the scattering photosphere, and thus greatly increasing the
range of wavenumbers that are damped by Compton drag off advected
radiation.

{\it Heavy Matter Loading and GRB Afterglow.}
One should also consider the effects of matter opacity in GRB outflows
with extremely high central compactness.  Although a core of the outflow
(e.g. near the rotation axis of the central engine) must attain very high
$\gamma_\infty \sim 100-300$, material off axis may not.\fonote{For example,
if the central engine is a rapidly-rotating neutron star or neutron torus,
then neutrino emission can easily power mass loss rates as high
as $(10^{51}~{\rm erg~s^{-1}}/c^2 \sim 10^{-3}$ g s$^{-1}$.}
Material expanding with $\gamma_\infty\sim 1-2$ becomes optically thin
to scattering on a timescale $\sim 1$ day $(E/10^{52}{~\rm erg})^{1/2}$.
This is comparable to the timescale on which the optical afterglow
detected from GRB970508 reached a maximum (Bond 1997).  This is telling,
since direct synchrotron emission between the contact discontinuity and
the forward shock probably is strongly suppressed due to the relative
weakness of the ambient magnetic field (Sect. 2.2).
A further motivation for {\it simultaneous} optical observations of GRBs
comes from the observation that the minimum frequency $N_c eB/m_ec 
\sim (10^2-10^3)\,eB/m_ec$ for optically thin cyclo-synchrotron emission lies
near $\sim 1$ eV (Sect. 3.3) near the $e^\pm$ scattering photosphere
and at a bulk Lorentz factor of $\sim 10^2$.  

{\it GRB Time Profiles: FRED vs. Chaotic.}
The light-curves of GRBs show a bewildering variety of shapes
(Meegan et al. 1996), but at
least two main classes can be identified:  bursts with smooth, asymmetric
pulses (`Fast Rise and Exponential Decay' or FRED); and more `chaotic' bursts
in which narrow and wide pulses are often superimposed and in which
the asymmetry of individual pulses is usually less clearly defined.  

Chaotic bursts are most easily explained if the dissipation in a burst
is driven by local physics within the outflow (PX94; T94; RM94; Sari \&
Piran 1997), rather than by interaction with an external medium.  This
leads to a simple discriminant between the two classes:  FRED bursts
arise from shells of ejecta that come into causal contact before
the $\gamma$-rays escape, and vice versa for chaotic bursts.  
However, if the FRED bursts are also powered by local physics on
a scale smaller than the width of the shell of ejecta, then the smoothness
of the lightcurves indicates that {\it the $\gamma$-ray emitting zone lies at
scattering depths $\tau_T >  1$}.  Given the lack of a clear spectral
distinction between the two classes, one reaches the same conclusion for
chaotic bursts.  Indeed, the transition zone between
large and small $\tau_T$ can be considerably broadened by pair creation,
and pairs are most effective at enhancing leptonic dissipative modes
near the scattering photosphere (Sect 3.3). In sum, this leads to the following
simple model:  a burst is smooth (FRED) or chaotic depending on
whether the {\it scattering photosphere} lies outside or inside the causal
contact radius $2\gamma_\infty^2 c\Delta t$.  

{\it Range of Temporal Frequencies and GRB Soft Tails.}
A flow will, in general, be variable on
a range of timescales $\Delta t$ and so dissipation can occur
over a range of optical depths.  The emergent spectrum varies 
considerably depending on whether most of the available energy
resides at long timescales or short.  

In this regard, it is interesting to note
that the extended soft bump following GRB 870303 (a chaotic burst) detected
by Ginga had a quasi-thermal cutoff (Yoshida et al. 1989), 
whereas the extended soft emission in GRB 960720 (a FRED burst) detected by 
BeppoSAX is closer to an extended powerlaw (with the possibility of a 
cutoff at $\sim 30$ keV; Piro et al. 1997).  This spectral difference could be
explained if the ejecta that produced the soft tail of GRB 870303
dissipated at $\tau_T^{e-p} \gg 1$ while they were causally disconnected
from the primary pulse of ejecta.  High energy cut-offs to GRB afterglow
are in general very diagnostic:  the high energy spectral index is
attracted to $\beta = -2$ below an energy $\sim (\gamma/\tau_T)m_ec^2$ (T94),
and so the presence of a high energy spectral break yields information
about a combination of bulk Lorentz factor and scattering depth (see also
Baring \& Harding 1997).

\begrefchapter{References}

\ref Baring, M.G. \& Harding, A.K. 1997, Ap. J. 481, L85

\ref Bisnovaty-Kogan, G.S. \& Lovelace, R.V.E. 1997, preprint

\ref Blandford, R.D. \& Payne, D.G. 1981, M.N.R.A.S., 194, 1041

\ref Blandford, R.D. \& Eichler, D. 1987,  Phys. Rep., 154 1

\ref Blandford, R.D. \& Levinson, A. 1995, Ap. J. 441, 79 (BL95)

\ref Bloemen, H., et al. 1995, A. A., 293, L1

\ref Blom, J.J., et al. 1995, A. A., 298, L33

\ref Bond, H.E. 1997, IAU circular 6654; see also circulars 6655-6658

\ref Cavallo, G. \& Rees, M.J. 1978, M.N.R.A.S., 183, 359

\ref Cohen, E., Katz, J.I., Piran, T., Sari, R., Preece, R.D., \&
Band, D.L. 1996, preprint.

\ref Dermer, C.D. \& and Schlickeiser, R. 1993, Ap. J., 416, 458

\ref Dermer, C.D., Miller, J.A., \& Li, H. 1996, Ap. J. 456, 106

\ref Done, C., Ghisellini, G., \& Fabian, A.C. 1990, M.N.R.A.S., 245, 1

\ref Duncan, R.C., Shapiro, S.L., \& Wasserman, I. 1986, Ap. J. 309, 141

\ref Duncan, R.C. \& Thompson, C. 1992, Ap. J. 392, L9

\ref Fatuzzo, M. \& Melia, F. 1996, Ap. J. 464, 316

\ref Fenimore, E.E., Klebesadel, R.W., \& Laros, J.G. 1996, 
Ap. J., 460, 964

\ref Ghisellini, G., Guilbert, P.W., \& Svensson, R. 1988, Ap. J., 334, L5

\ref Haardt, F. \& Maraschi, L. 1993, Ap. J. 413, 680

\ref Hoshino, M. \& Arons, J. 1991, Phys. Fluids B, 3, 818

\ref Hurley, K., et al. 1994, Nat., 372, 652

\ref Jaroszynski, M. 1993, A. A., 43, 183

\ref Levinson A. 1996, Ap. J., 459, 520

\ref Lin, R.P. 1990, in {\it Basic Plasma Processes on the the Sun},
ed. E.R. Priest \& V. Krishan, p. 467

\ref Mahadevan, R., Narayan, R. \& Yi, I. 1996, Ap. J. 465, 327

\ref Mallozzi, et al. 1995, Ap. J. 454, 597

\ref Mannheim, K. 1993, A. A. 269, 67

\ref Maraschi, Ghisellini, \& Celotti, 1992, Ap. J. 397, L5

\ref Marcowith, A., Henri, G., \& Pelletier, G. 1995, M.N.R.A.S., 277, 681

\ref McNaron-Brown, et al. 1995, Ap. J. 451, 575

\ref Meegan, C.A., et al. 1996, Ap. J. 106, 65

\ref Melrose, D.B. \& Pope, M.H. 1993, Proc. Ast. Soc. Aust., 10, 222

\ref M\'esz\'aros, P., Rees, M.H. \& Papathanassiou, H. 1994, Ap. J. 432, 181
(MRP94) 

\ref M\'esz\'aros, P. \& Rees, M.J. 1997, Ap. J. 482, 29

\ref Paczy\'nski, B. 1990, Ap. J., 363, 218

\ref Paczy\'nski, B. \& Xu, G. 1994, Ap. J. 427, 708 (PX94)

\ref Papathanassiou, H. \& M\'eszar\'os, P. 1996, Ap. J. 471, L91

\ref Pendleton, G.N., et al., 1996, in proceedings of the
Third Huntsville Symposium on Gamma-Ray Bursts, ed. C. Kouveliotou, M.S. Briggs
\& G.J. Fishman, p. 228

\ref Piran, T. \& Shemi, A. 1993, Ap. J. 403, L67

\ref Piran, T. \& Narayan, R. 1996, in proceedings of the
Third Huntsville Symposium on Gamma-Ray Bursts, ed. C. Kouveliotou, M.S. Briggs
\& G.J. Fishman, p. 233

\ref Piro, L., et al. 1997, preprint

\ref Rees, M.J. \& M\'esz\'aros, P. 1992, M.N.R.A.S., 258, 41

\ref Rees, M.J. \& M\'esz\'aros, P. 1994, Ap. J. 430, 93 (RM94)

\ref Romanova, M.M. \& Lovelace, R.V.E. 1992, A. A., 262, 26

\ref Sari, R. \& Piran, T. 1997, Ap. J., 485, 270

\ref Shapiro, S.L., Lightman, A.P., \& Eardley, D.M. 1976, Ap. J., 204, 187

\ref Sikora, M.,  Begelman, M.C., \& Rees, M.J. 1994, Ap. J. 421, 153

\ref Sikora, M., Madejski, G., Moderski, R., \& Poutanen, J. 1997, Ap. J.
484, 108

\ref Stern, B., Poutanen, J., Svensson, R., Sikora, M., \& Begelman, M.C.
1995, 449, L13

\ref Thompson, C. 1994, M.N.R.A.S., 270, 480 (T94)

\ref Thompson, C. \& Duncan R.C. 1995, M.N.R.A.S., 275, 255

\ref Thompson, C. 1996, in proceedings of the
Third Huntsville Symposium on Gamma-Ray Bursts, ed. C. Kouveliotou, M.S. Briggs
\& G.J. Fishman, p. 802 (T96)

\ref Thompson, C. \& Blaes, O. 1997, submitted to Phys. Rev. D (TB97)

\ref Usov, V.V. 1992, Nat., 357, 472

\ref Usov, V.V. 1994, M.N.R.A.S., 267, 1035

\ref Vietri, M. 1996, Ap. J. 471, L95

\ref Yoshida, A., et al. 1989, P.A.S.P., 41, 509 

\endref

\byebye

\documentstyle[psfig,fleqn]{rel}
\textheight 190.0mm
\textwidth 130.0mm
\input macro.tex
\begin{document}
\pagenumbering{arabic}
\addtocounter{page}{62}
\frenchspacing 
\parindent15pt 
\oddsidemargin 0mm
\evensidemargin 0mm

\renewcommand{\theequation}{\arabic{equation}}
\setcounter{equation}{0}
\pagestyle{plain}

\leftline{}
\leftline{}
\leftline{}
\leftline{}

\Large
\centerline{\bf Dissipation in Relativistic Outflows:}
\centerline{\bf A Multisource Overview}

\normalsize 
\leftline{}
\leftline{}

\centerline{\large Christopher Thompson}

\leftline{}

\small 
\centerline{Physics and Astronomy, University of North Carolina,}
\centerline{Chapel Hill, NC 27599}

\vspace{18mm}

\footnotesize 
\noindent 
{\bf Abstract.} Relativistically expanding sources of X-rays and $\gamma$-rays
      cover an enormous range of (central) compactness and Lorentz factor.
      The underlying physics is 
      discussed, with an emphasis on how the dominant dissipative mode
      and the emergent spectrum depend on these parameters.  
      Photons advected outward from high optical depth are a potentially
      important source of Compton seeds.  Their characteristic energy
      is bounded below by $\sim 1$ MeV in pair-loaded outflows of
      relatively low compactness, and remains near $\sim 1$ MeV at
      very high compactness and low matter loading.  This is compared
      with the characteristic energy of $O(1)$ MeV observed in the
      rest frame spectra of many sources, including $\gamma$-ray bursts,
      OSSE jet sources, MeV blazars, and the intense initial 0.1 s pulse
      of the March 5 event.  Additional topics discussed include the
      feedback of pair creation on  electron heating and the formation of
      non-thermal spectra, their effectiveness at shielding the dissipative
      zone from ambient photons, direct Compton damping of irregularities
      in the outflow, the relative importance of various soft photon sources,
      and the softening of the emergent spectrum that results from heavy
      matter loading.  The implications of this  work for X-ray and optical
      afterglow from GRB's are briefly considered.  Direct synchrotron
      emission behind the forward shock is inhibited by the
      extremely low energy density of the ambient
      magnetic field.   Mildly relativistic ejecta off axis from the
      main $\gamma$-ray emitting cone become optically thin to scattering
      on a timescale of $\sim 1$ day $(E/10^{52}~{\rm erg})^{1/2}$, and
      can be a direct source of afterglow radiation.

\vspace{15mm}
\normalsize

\leftline{\large\bf 1. Introduction:  variety of sources and spectral behavior}

\vspace{4mm}

\noindent 
X-ray and $\gamma$-ray emission from relativistic outflows is
powered by the conversion of bulk kinetic energy and Poynting luminosity,
by a variety of possible mechanisms.  However, the assumed values of
key parameters such as the Lorentz factor $\gamma$
of the outflow, the compactness of the central source, as well
as the optical depth and size of the
dissipative zone, vary dramatically between the different classes of sources.
For example, $\ell_c \sim 10^{15}$ and $\gamma\sim 10^2{-}10^3$ 
are inferred for cosmological
$\gamma$-ray burst (GRB) sources vs. $\ell_c \sim 10{-}10^2$ and $\gamma 
\sim 3{-}30$ for blazars (Fig.~1).
This motivates a more global analysis of how dissipation of kinetic and
magnetic energy is achieved, which provides some interesting new perspectives
on particular sources.   
\eject

Another key point is that sources (or types of sources) exhibit
different {\it spectral states}.  GRBs are predominantly
non-thermal, but some 
contain subluminous precursors, tails, and sub-pulses with 
distinctly thermal high energy cutoffs (Yoshida et al. 1989;
Pendleton et al. 1996).  Blazars are occasionally observed
with emission peaked strongly at $\sim 1$ MeV, in distinction to the
more usual extended power-law behavior 
(Bloemen et al. 1995; Blom et al. 1995). And the remarkable 5 March 1979
burst was initiated by an extremely intense $\sim 0.1$ sec flare whose
luminosity exceeded that of the remainder of the burst by a factor
$\sim 300$ and showed much more pronounced spectral softening
(Fenimore et al. 1996).  

\pagestyle{headings}
\markboth{Thompson}
{Dissipation in Relativistic Outflows: A Multisource Overview}

\begin{figure}[t]
\centerline{\psfig{file=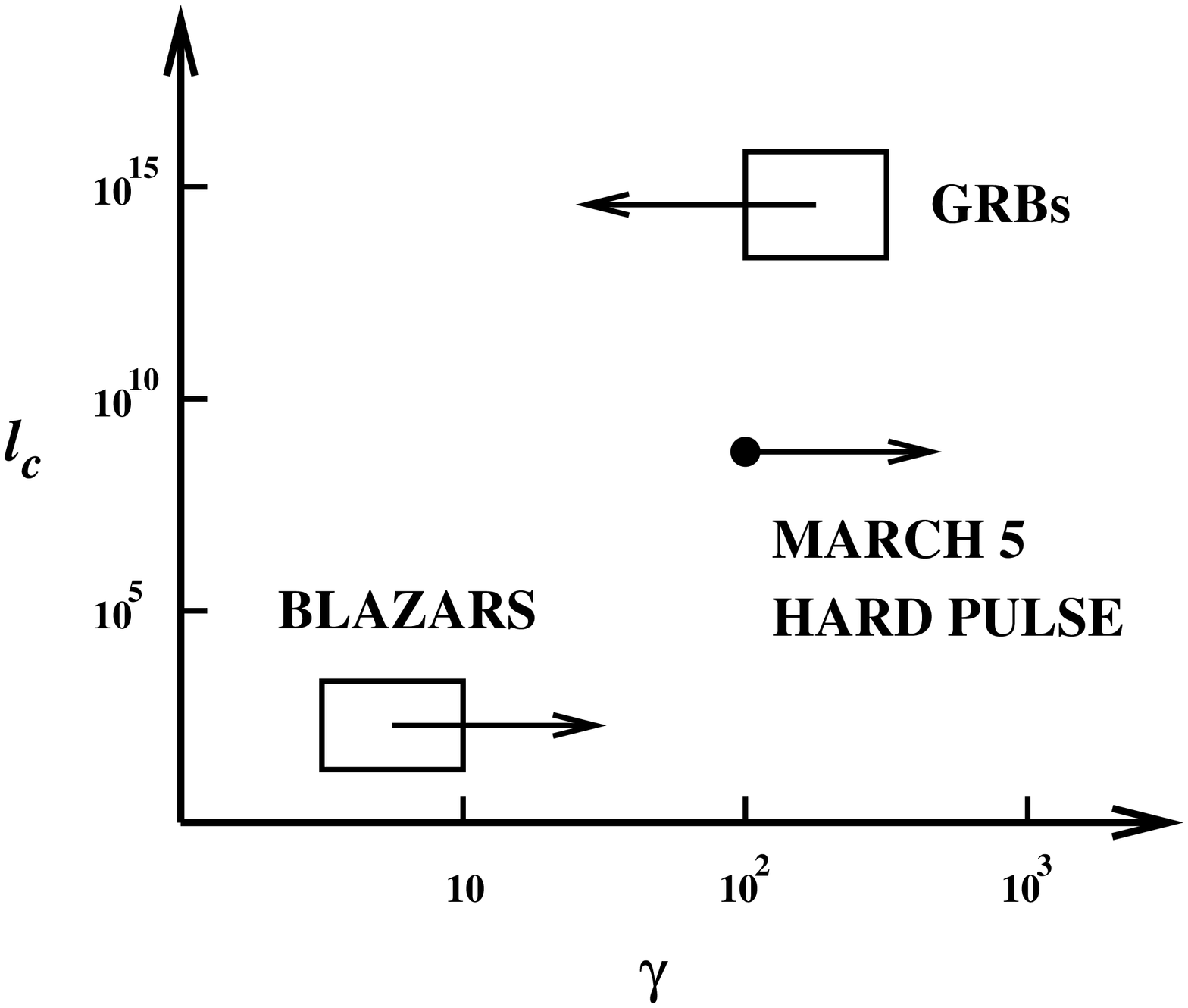,width=7cm,angle=90}}
\vspace*{4mm}
\caption{Blazars, $\gamma$-ray burst sources and
the initial 0.1 sec hard pulse of the March 5, 1979 burst occupy
distinct regions of the plane defined by source compactness $\ell_c$
and asymptotic Lorentz factor $\gamma_\infty$.  GRB outflows may contain
lower-$\gamma_\infty$ ejecta that manifests itself as softer
tails, precursors and sub-pulses (T94, T96).  
The Lorentz factors of blazar sources
are only indirectly constrained by superluminal motion outside the
$\gamma$-ray emitting region (cf. Wagner, these proceedings).}
\vspace*{-2mm}
\end{figure}

Spectral states with sharp high energy cutoffs have a simple interpretation
as the residue of an {\it optically thick\/} outflow.  Indeed, at very high
$\ell_c$, the outflow is self-shielding from the central radiation source,
as well as from external (e.g. side-scattered) radiation.   
Radiation advected outward from large scattering depth then becomes an
important source of seeds for extended non-thermal spectra --- 
in addition to the more familiar optically thin synchrotron-self-Compton
mechanism.  This means that the inner boundary conditions on the flow
are much more important than is usually supposed.   And this raises
an interesting question:  are sources of relatively low compactness
(e.g. blazars) ever self-shielding in this manner?

It is intriguing to note, in this regard, that the characteristic energy
$\sim 1$ MeV appears in a number of sources (of widely varying parameters):
the spectral breaks observed in OSSE jet sources ($\sim 1{-}3$ MeV after
compensating for cosmological redshift; e.g. McNaron-Brown et al. 1995)
and in classical GRBs ($\sim 100$ keV $-1$ MeV at peak luminosity, 
without compensating for redshift; Mallozzi et al. 1995);  in MeV blazars;
as well as the initial hard spike of the March 5 burst ($\sim 300$ keV).
Of course, the possibility that selection effects narrow the
observed spectral break distribution should be considered 
carefully.  This happens in the case of the GRB sources 
(Piran \& Narayan 1996) only if the total burst energy is constrained to
yield an inverse correlation between break energy and flux --- in distinction
to the strong positive correction observed within individual bursts.
Most plausible cosmological GRB sources release as much as $10^{53}{-}10^{54}$
erg, which allows for a fraction of very energetic and hard bursts.

After careful consideration of various photon sources, it turns out that
an advected Wien peak maintains a characteristic energy of $\sim 1$ MeV 
over a wide range of central compactness, if the flow is sufficiently
relativistic (Sects. 1.1; 3.1).   Pair creation
by photon collisions, $\gamma + \gamma \rightarrow e^+ + e^-$,
which has traditionally been viewed as inimical to high energy gamma-ray
production, can in fact play an essential role by i) increasing the
efficiency of leptonic dissipative modes; ii) reducing the lower cut-off energy
of the non-thermal pair distribution to $\gamma_{\rm min} \sim 1$, 
which yields a break in the spectral distribution of Compton-upscattered
Wien photons near the position of the original Wien peak; and iii)
selecting non-thermal over thermal spectra at Comptonizing hotspots
a high-$\gamma$ outflow (Sect. 3.3).  
Since the photon collision cross section is
comparable to Thomson, feedback from pair creation works most effectively
near $\tau_T \sim 1$.  This contrasts with the inhomogeneous
external Comptonization model (Blandford and Levinson 1995, hereafter
BL95), where the non-thermal high energy continuum emerges well outside
the scattering photosphere.  Indeed, the position of the scattering
photosphere in a pair-loaded outflow is sensitive to the amount of
continuous heating, and pair creation can significantly broaden
the transition zone between optically thick and thin flows (Sects. 1.1, 2.2).

Thus, by focussing on those aspects of the physics that are special to the
large-$\ell_c$ GRB regime, and then considering how these vary with
compactness, interesting new insights can be obtained on both the
GRB and blazar problems.  When constructing GRB models, it is sobering to
realize that blazars are still far from being understood, even with the
much broader spectral information available.

In these notes, I will explore the following additional points:
\smallskip 

\litem{$\bullet$}  The dependence of the dominant dissipative mode on optical depth.
 At $\tau_T > 1$ direct
Comptonization by bulk fluid motions is most effective; whereas 
non-thermal (e.g. Fermi) particle acceleration is a crucial
ingredient of any radiative model at low $\tau_T$.  
Strong-wave acceleration can be excluded if enough scattering charges
are present to generate an observable flux of Comptonized
high energy photons.

\litem{$\bullet$}  What{\kern-.05em} is{\kern-.05em}
the{\kern-.05em} relative{\kern-.05em} importance{\kern-.05em}
of{\kern-.05em} double{\kern-.05em} Compton{\kern-.05em}
emission{\kern-.05em} and{\kern-.05em} cyclo-synchrotron emission as seeds for Comptonization (at large $\ell_c$)?

\litem{$\bullet$} How does the influence of geometrical effects (e.g. beaming)
vary with $\gamma_\infty$?

\litem{$\bullet$} High energy cutoffs to extended power-law spectra are
extremely diagnostic.  In GRB sources these are very poorly constrained.
Measurements by EGRET in the 30 MeV${-}$10 GeV range indicate
a significant decorrelation with the 1 MeV flux, with\, the\, high\, energy\,
emission\, often\, being\, significantly\, {\it delayed\/}\, (Hurley\break et al. 1994).

\vspace{5mm}
\leftline{\bf 1.1. The $\ell_c$-$\gamma_\infty$ Plane}
\vspace{3mm}

\noindent 
The variety of possible dissipative modes is neatly summarized in
a two-dimensional plane labeled by (Fig.~2)
	\begin{equation}
\ell_c = {L_{\rm rel}\sigma_T\over 4\pi m_ec^3 R_0}~~~\gamma_\infty = 
{L_{\rm rel}\over \dot Mc^2}.
	\end{equation}
The radius $R_0$ of the central engine is identified with the
Alfv\'en radius or light-cylinder radius as appropriate.
At risk of oversimplification I will usually assume that
$\gamma$ has attained the limiting value $\gamma_\infty = 
L_{\rm rel}/\dot Mc^2$ due to matter loading $\dot M$ (when discussing
delayed dissipation at large distances from the central engine).

\begin{figure}[h]
\vspace*{2mm}
\centerline{\hspace*{-3mm}\psfig{file=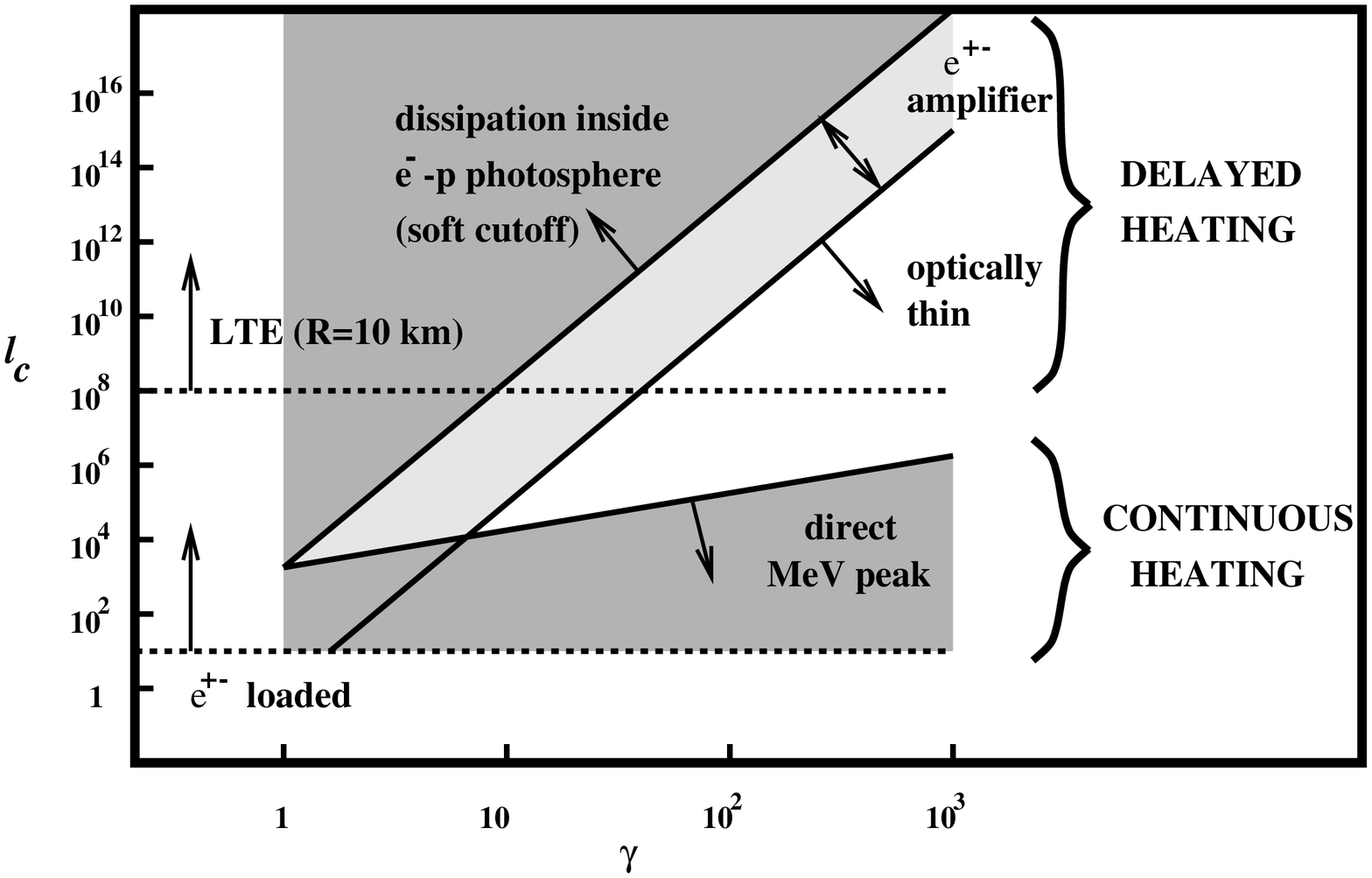,height=6.cm,angle=90}}
\vspace*{4mm}
\caption{The variety of dissipative regimes in relativistic outflows
as summarized in the $\ell_c$-$\gamma_\infty$ plane.  Flows
with heavy matter loading and asymptotic Lorentz factor 
$\gamma_\infty < \gamma_{e-p}$ occupy the top left region.  They are
a plausible source of soft subcomponents of GRBs.
Flows with $\gamma_{e-p} < \gamma_\infty <  \gamma_{e^\pm}$ can become
pair loaded, and occupy the adjoining stripe.  Flows with low
matter loading, $\gamma_\infty > \gamma_{e^\pm}$, dissipate at low
optical depth.   Above the upper horizontal dashed line, 
local thermodynamic equilibrium is achieved at the base of the flow;  
whereas below this line the photon distribution is Wien.
All of the flows in this region are assumed to 
undergo {\it delayed\/} dissipation after thermal pairs freeze out,
at a radius where inhomogeneities (such as reconnecting magnetic
fields and shocks) regain causal contact.   By contrast, flows in the
lower portion of the diagram are assumed to dissipate {\it continuously\/}
and remain pair loaded out to the scattering photosphere.   This is
expected in collimated flows with lower asymptotic Lorentz factors (such
as blazars).  Large scattering depths cannot be maintained below the lower
horizontal dashed line.}
\end{figure}

The radius at which dissipation takes place is limited significantly
by causality at large $\gamma$, as long as 
$\gamma(r)$ grows faster with radius than $r^{1/2}$
near the base of the outflow.  Thus, it is convenient to define the
compactness $\ell_{\Delta t} = \ell_c\times (R_0/c\Delta t)$ associated with
variability on a timescale $\Delta t$.   A plausible value of the dissipative
radius is $\sim 2\gamma_\infty^2 c\Delta t$
for a variety of dissipative modes, including magnetic reconnection and
MHD turbulence (Romanova \& Lovelace 1992, hereafter RL92; 
Thompson 1994, 1996, hereafter T94, T96; Levinson, these proceedings) 
and shocks powered by variations in $\dot M$ (Pa\-czy\'n\-ski \& Xu 1994, hereafter
PX94; Rees and M\'esz\'aros 1994, hereafter RM94).
The radius of the scattering photosphere of the wind is related to $R_{\rm diss}$
by a very strong function of $\gamma_\infty$,
	\begin{equation}
{R^{e-p}_{\tau=1}\over 2\gamma_\infty^2 c\Delta t} \sim 
{m_e\over m_p}{\ell_{\Delta t}\over \gamma_\infty^5}~~~(n_p \gg 
n_{e^+}),
	\end{equation}
when the scattering depth is dominated by the advected electron-ion
contaminant.  A~non-thermal photon tail extending above energy $m_ec^2$
in the rest frame will greatly increase the density of scatterers, with
the result
	\begin{equation}
{R^{e^\pm}_{\tau=1}\over 2\gamma_\infty^2 c\Delta t} \sim 
{\ell_{\Delta t}\over \gamma_\infty^5}~~~(n_{e^+} \gg n_p)
	\end{equation}
for a photon index $\beta \sim -2$ characteristic of GRBs.  The effects
of pairs are discussed further in Sect. 3.

Outflows with Lorentz factor $\gamma_\infty < \gamma_{e-p} = 
(m_e/m_p)^{1/5}\ell_{\Delta t}^{1/5}$ dissipate well inside the
electron-ion photosphere.  They occupy the upper left portion of
Fig.~1, and have a~possible association with the soft X-ray precursors,
tails and quasi-thermal sub-pulses of GRBs (T96).    
Outflows with $\gamma_{e-p} < \gamma_\infty 
< \gamma_{e^\pm} = \ell_{\Delta t}^{1/5}$
dissipate inside the the pair photosphere, if the high energy continuum
extends up to $\sim m_ec^2$ in the local rest frame of the flow.
And, outflows with $\gamma_\infty > \gamma_{e^\pm}$
dissipate at low scattering depth, independent of the efficiency of
pair creation.  
The spectral consequences of variations in the matter
loading are discussed further in Sect. 3.

Flows in which the irregularities maintain causal contact 
will undergo continuous heating while optically thick.  An interesting 
example of such a flow is the low-$\gamma$ sheath of a 
relativistic jet, in which the luminosity of entrained photons increases
with radius as the kinetic energy of the higher-$\gamma$ core of the jet
is dissipated.  Although 
the relation between scattering depth and internal temperature
of the flow is sensitive to the high energy distributions of the pairs
and photons,  the mean energy of the emergent photon spectrum is regulated to
near $\sim\gamma m_ec^2$.  In the case where the photon distribution is Wien, 
the strong $T$-dependence 
$n_{e^\pm}/n_\gamma = (\pi/2)^{1/2}(T/m_ec^2)^{-3/2}\exp(-m_ec^2/T)$
of the equilibrium pair density  guarantees that as soon as $T$ drops
much below $m_ec^2$ in the rest frame, the flow becomes optically thin.
This yields a simple relation between the observed (Lorentz-boosted) 
temperature and the temperature $T_0$ at the base of the flow,
	\begin{equation}
	T_{\rm obs} = T_0\,(L_{\gamma}/L_{\gamma\,0}),
	\end{equation}
in terms of the growth of the photon luminosity from $L_{\gamma\,0}$
to $L_\gamma$.  

The pair 
density resulting from direct heating of the photons by MHD turbulence 
can be less sensitive to $T$.  I assume that the bulk of the
photon energy lies in a Wien bump, but that wave energy is
excited in transient surges (e.g. by reconnection) with an equivalent
temperature $T_w$ somewhat in excess of $m_ec^2$.  Then a significant
fraction of the wave energy is converted to photons above the pair
production threshold.  Balancing this source against pair annihilation,
the scattering depth in the direction perpendicular to a jet (of opening
angle $\theta$) is $\tau_{T\perp} = {1\over 2}n_e\sigma R\theta \sim 
(\gamma\ell_\gamma)^{1/2}$, where
$\ell_\gamma$ is the compactness (1) in the advected photons at radius $R$.
At $\tau_T > 1$ ($\ell_\gamma > 1$), 
freshly created pairs Compton cool before annihilating (the cooling
timescale being shorter by a factor $\sim \tau_{T\parallel}^{-1}$)
and carry a fraction $\sim \tau_{T\parallel}^{-1}$ of the total energy of
the flow.  
This regulates the Compton parameter induced by mildly relativistic pairs
to $y \simeq \tau_{T\parallel}^{-1}\cdot\tau_{T\parallel} \sim 1$.  The
annihilation photons Compton downscatter off the cooled pairs
to an energy $\sim m_ec^2/\tau_{T\parallel}$ in the rest frame. 
The photon spectrum emerging at the pair photosphere then peaks
at an energy $\sim \gamma m_ec^2$.
This implies only a modest increase in the mean energy per photon
along the jet, by a factor $\sim \gamma m_ec^2/3T_0$,
which is easily supplied from the high-$\gamma$ core to the low-$\gamma$
sheath.

Continuous heating by relativistic electrons at low optical depth has
been considered by Sikora et al. (1997) as a model for the MeV blazars.  
They relate the peak energy to the observed variability timescales, but
since this energy is not directly tied to a microphysical scale, it could
be expected to lie well below $\sim 1$ MeV in some sources.
By contrast, direct Compton damping of mildly relativistic turbulence
in an optically thick jet yields a Wien peak energy that is bounded below
by $\sim 1$ MeV (although still dependent on bulk $\gamma$ and the
energy transferred from the bulk motion to the photons).

\vspace{8.5mm}
\leftline{\large\bf 2. Dissipative mechanisms}
\vspace{6mm}

\leftline{\bf 2.1. Sources of free energy}
\vspace{4.5mm}

\noindent 
The internal sources of free energy in a relativistic flow can be broadly 
divided into two categories:  those associated with radial and angular
inhomogeneities.  Both appear to be important in jet sources, and
while both have also been considered in GRB sources, radial inhomogeneities are
probably more important in the large-$\gamma$ \mbox{context.}  Reconnection surfaces
and variations in the ratio of particle pressure to magnetic pressure 
will lead to internal heating of strongly-magnetized outflows
(RL92; T94), as will kinetic energy
fluctuations in particle-dominated outflows (PX94; RM94).
However, interactions with an external medium (Rees and 
M\'esz\'aros 1992) can become significantly non-spherical as the outflow
in a $\gamma$-ray burst source decelerates (Sect. 4).

Which of these energy sources dominates depends on the strength of the
magnetic field in the outflow and the radius of the dissipative zone.
It appears that $\gamma_\infty \sim 100{-}300$
can be achieved in an outflow of luminosity $\sim 10^{51}$ erg s$^{-1}$
only if it is Poynting-flux dominated at the
source.  Indeed, {\it any\/} triggering scenario for a GRB that produces
an object with the density of nuclear matter and a rotation period of
$\sim 10^{-3}$ s plausibly involves magnetic fields as strong as
$\sim 10^{15}$ G through dynamo amplification and thus leads to 
a rotationally-driven
MHD outflow of luminosity $L_P \sim 10^{50}{-}10^{51}$ erg s$^{-1}$ (T94;
see also Duncan \& Thompson 1992; 
Usov 1992, 1994; Vietri 1996; M\'esz\'aros and Rees 1997
for particular models).  The alternative mechanism of
$\nu-\bar\nu$ annihilation into $e^\pm$ pairs has an efficiency of
$\sim 10^{-3}$ (Jaroszynski 1993), and so the neutrino luminosity
required to power a $\gamma$-ray flux of $\sim 10^{51}$ erg s$^{-1}$
drives a~mass flux (by absorption on nucleons) that exceeds the tolerable
value by $\sim 10^6$ (cf. Duncan, Shapiro and Wasserman 1986).  
\eject 

This advected magnetic field almost certainly has an important
effect on the dissipative mechanism.
Quasi-perpendicular shocks are significantly weakened even when
the magnetic field carries a few percent of the energy flux;  this
in turn \mbox{steepends} non-thermal particle spectra arising from
first-order Fermi acceleration.   Although particle acceleration at
oblique relativistic shocks can be quite efficient (as discussed by
Kirk, these proceedings), one expects that {\it internal\/} shocks in
GRB outflows with $\gamma \sim 100{-}300$ are approximately radial.
Thus, GRB models based on internal shocks may unfortunately require
complicated departures from spherical symmetry.

\vspace{5mm}
\leftline{\bf 2.2. Geometrical effects:  pair cocoons}
\vspace{3mm}

\noindent 
The huge compactness of a GRB source ($\sim 10^{15}$) causes
it to be {\it self-shielding}.  Not only is the central engine
hidden from the dissipative zone, but the optical depth
$\tau_\perp$ perpendicular to the axis of the flow exceeds the parallel
depth by the very large factor,
	\begin{equation}
{\tau_\perp\over\tau_\parallel} \sim \gamma^2\theta \sim 10^{4-5}\theta,
	\end{equation}
for reasonable values of the opening angle $\theta$.

\begin{figure}[h]
\vspace*{2mm}
\centerline{
	\psfig{file=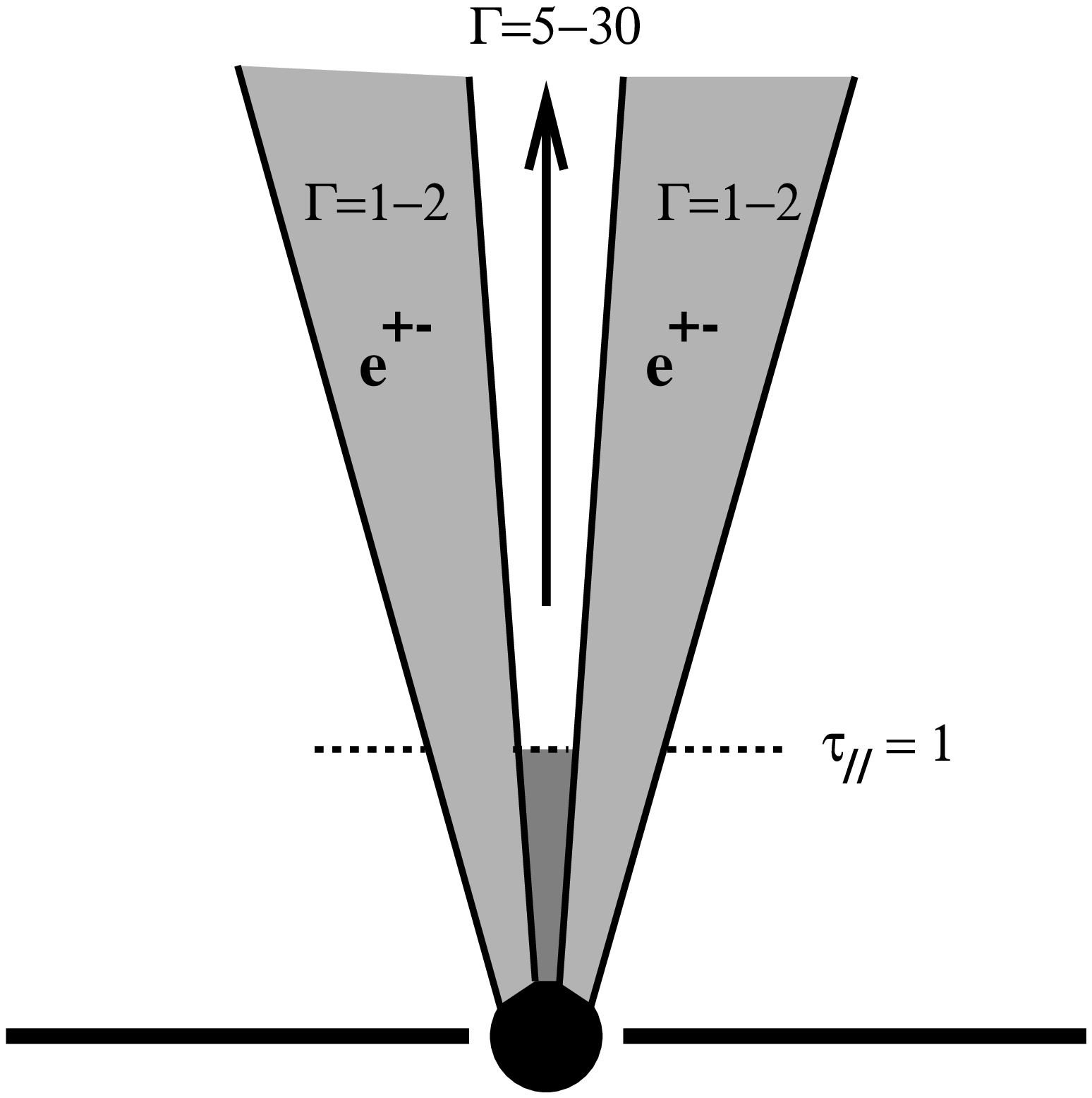,height=6.cm,angle=90}}
\vspace*{4mm}
\caption{A sheath of low-$\gamma$ material surrounding the
high-$\gamma$ core of a jet can remain pair loaded out to a large
distance (6) from the central engine.  The scattering photosphere
of this {\it pair cocoon\/} will in general lie far outside the radius
at which the core becomes transparent to high energy $\gamma$-rays.}
\end{figure}

This raises the question:  as $\ell_c$ decreases toward the values 
typical of AGN, at what point does the central engine become visible?
One intriguing possibility is that the engine remains shielded in some blazars,
with the high-$\gamma$ cores of the jet being surrounded by a 
{\it pair cocoon\/} (Fig.~3). Because Lorentz dilation causes $\tau_\parallel$ to
decrease in proportion to $\gamma^{-2}(L_{\rm jet}/\dot M_{\rm jet} c^2)^{-1} 
\propto \gamma^{-3}$, the
high-$\gamma$ core of a~jet can become optically thin along its axis
well inside the photosphere of the $\gamma \sim 1{-}2$ sheath.  This
lies at a~radius
	\begin{equation}
R_{\tau= 1}^{e^\pm} \sim 2\times 10^{18}\,\left({L_{\rm jet}\over
10^{47}~{\rm erg~s^{-1}}}\right)\,\left({\theta\over
0.1}\right)^{-1} {\rm cm},
	\end{equation}
assuming that the sheath acquires a non-negligible fraction of the kinetic
energy at this radius.    All that
is required (Sect. 1) is that the sheath be i) pair-loaded and optically
thick at its base, and ii) {\it continuously\/} heated (e.g. by Kelvin-Helmholtz
instabilities with higher-$\gamma$ material) at a sufficient rate
to overcome the effects of adiabatic cooling.  Condition ii) is
plausibly satisfied in the cores of superluminal sources, and
i) is satisfied if the outflow is strongly turbulent at its base (Sect. 3).  
Below this threshold heating rate,  the position of the pair photosphere
is very sensitive to the amount of heating.

This behavior is quite different from what may occur in the
boundary layer between a sub-relativistic, collimated jet and an
optically thick accretion torus supported by radiation pressure.
Arav and Begelman (1993) argue that, in the absence of pairs,
the matter density and scattering
depth are in fact {\it suppressed} in such a boundary layer
with respect to both the accretion torus and jet core.

\vspace{2mm}
\leftline{\bf 2.3. Diverse photon sources}
\vspace{3mm}

\noindent 
When the flow is self-shielding in this manner, {\it advected\/} 
quasi-thermal radiation
becomes an important source of Compton seeds (T94, T96).
Even though external radiation exerts a stronger drag force
than internally generated radiation by a factor $\gamma^2$
(Sikora, Begelman, \& Rees 1994; BL95), it
cannot penetrate the high-$\gamma$ component of the flow.
This also disfavors models involving acceleration
and heating of the jet by central continuum radiation propagating
along the jet axis (Dermer and Schlieckeiser 1993; 
Marcowith et al. 1995).
The main competing source of seed photons is then synchrotron
radiation (e.g. Maraschi et al. 1992; M\'esz\'aros, Rees and 
Papathanassiou 1994, hereafter MRP94).

In quantifying the relative importance of these two radiation
sources, one must consider separately thermal and non-thermal scatterers.
The advected radiation characteristically has a much higher frequency
than the cyclotron frequency, and if its energy density $U_\gamma^a$
is comparable to $B^2/8\pi$, then it will dominate the
magnetic field as a coolant of thermal electrons, due to self-absorption
near the cyclotron resonance.   Parametrizing the frequency at
which the cyclo-synchrotron radiation becomes self-absorbed in terms
of a cyclotron harmonic $N_{sa}$, one deduces from Kirchoff's law that
C-S cooling rate is smaller than the Compton cooling
rate off the advected radiation by the factor
	\begin{equation}
{\nu j_\nu \over 4\sigma_T n_e c(T/m_ec^2)
U_\gamma} = {2\alpha_{em}\over\pi} {N_{sa}^3\over \tau_T}\,
\left({B\over B_{QED}}\right)\,{B^2/8\pi\over
U_\gamma},~~~\left(\nu = N_{sa} {eB\over 2\pi m_e
c}\right)
	\end{equation}
where $\alpha_{em} = 1/137$ and $B_{QED} = 4.4\times 10^{13}$ G.  For
example, one expects $B/B_{QED} < 10^{-8}$  
near the scattering photospheres of GRB outflows, so that
C-S cooling can be neglected even if $N_{sa} \sim 300$. 

Another effect of the advected radiation is to limit the Compton
parameter\break $dy/d\ln R \simeq 4(T/m_ec^2)(1+4T/m_ec^2)d\tau_T/d\ln R$ in 
optically thick regions of the flow.  The magnitude of this effect
depends on the mean energy per photon $\langle h\nu\rangle$, 
and the bulk Lorentz factor $\gamma$.  If the flow is photon rich,
with $\langle h\nu\rangle/\gamma \ll m_ec^2$ in its rest frame,
then the advected photons limit $y$ to a value $\sim \ln(\langle \nu\rangle/
\nu_0)$, where $\nu_0$ is peak frequency before the advected photons
undergo (delayed) reheating.  This prevents very low energy C-S photons
from being upscattered much in frequency.  A stronger limit
$y \sim 1$ applies in outflows that are continuously Compton heated by
MHD waves and shocks (Sect. 1.1).   Nonetheless,
a much larger $y$-parameter will be maintained at the
base of the outflow, where the photon flux has a net divergence (Sect. 3.1).

Synchrotron radiation from blazar sources is usually ascribed to
the same high energy electrons/positrons that are responsible for the
X-ray/$\gamma$-ray emission.  The high energy cutoff of the 
synchrotron peak has been ascribed to a suppression of the non-thermal
particle density inside the scattering photosphere (e.g. Levinson 1996),
but mildly relativistic pairs are only suppressed by a factor
$\sim \tau_T^{-1}$.  Thus, the particle spectrum must itself
steepen considerably at $\tau_T > 1$.
Rather large minimum non-thermal Lorentz factors $\gamma_{\rm min}$ are sometimes
conjectured in GRB sources, sufficient to place the synchrotron
peak energy directly in the MeV range (e.g. MRP94).
However, synchrotron absorption becomes much
more important in that context, if the outflow
becomes sufficiently pair loaded that $\gamma_{\rm min} \sim 1$ (Sect. 3.3).

Advected{\kern-.05em} radiation{\kern-.05em} can{\kern-.05em}
also{\kern-.05em} be{\kern-.05em} the{\kern-.05em}
dominant{\kern-.05em} coolant{\kern-.05em} at{\kern-.05em} {\it external\/}{\kern-.05em}
relativistic{\kern-.05em} shocks.  Consider a shell of relativistic matter of
initial width $c\Delta t$.  After the shell becomes optically thin
to scattering, the radiation moves ahead of the shell, but continues
to overlap inside a radius $R > 2\gamma^2\,c\Delta t$.  The point is
that when the external medium is cold (with a sound
or Alfv\'en speed much less than $c$), the external magnetic field $B_{ex}$
that is swept up and compressed in between the forward shock and
the contact discontinuity typically has a {\it much\/} smaller energy
density than the radiation, by a~factor
	\begin{equation}
{B^2/8\pi\over U_\gamma} \sim 2\times 10^{-10}
\left({B_{ex}\over 3\times 10^{-6}{\rm ~G}}\right)^2
\left({L_\gamma\over 10^{50}~{\rm erg~s^{-1}}}\right)^{-1}\,
\left({\gamma\over 10^2}\right)^8\left({\Delta t\over 10~{\rm s}}\right)^2
	\end{equation}
at $R = 2\gamma^2c\Delta t$ (assuming a compression factor of 7).
{\it This strongly suggests that direct synchrotron radiation from 
the forward shock is {\rm not} the mechanism primarily
responsible for delayed optical and X-ray emission from GRBs.}
Direct radiation from a low-$\gamma$ component of the ejecta is
considered in Sect. 4.

\vspace{5mm}
\leftline{\bf 2.4. Dissipation at $\tau_T > 1$}
\vspace{3mm}

\noindent 
MHD waves, turbulent motions and shocks can in principle tap a
significant fraction of the energy of the outflow.  A periodic excitation of
frequency $\omega_k$ involving a~displacement $\xi_k$ of the fluid
deposits its energy {\it directly\/} in the photons via Compton drag 
when the scattering depth across $\xi_k$ is less than $\sim c/\omega_k\xi$
(T94).  For example, relativistic Alfv\'en waves have a damping time 
	\begin{equation}
	t_{\rm drag} \omega_k \sim {(\delta B)^2_k/8\pi\over U_\gamma},
	\end{equation}
which is shorter than the wave period as long as the photon gas has
a higher energy density than waves {\it in the relevant range of wavenumbers\/}
(Thompson and Blaes 1997, hereafter TB97).
This mechanism is particularly effective near the scattering photosphere,
and is also effective even at large $\tau_T$,
as long as higher wavenumber turbulence is generated via a turbulent
cascade.
The wave amplitude generally decreases with wavenumber in such a cascade,
with the result that $(\delta B)_k^2/8\pi \ll U_\gamma$ at dissipative
scales even if $(\delta B)_k^2/8\pi \sim U_\gamma$ at the outer scale.
Shocks also transfer energy to the photon fluid via compression and direct
first-order Fermi acceleration (Blandford and Payne 1981), 
although energy transfer is slowed significantly when the photon pressure
becomes comparable to the material ram pressure ahead of the shock.

\vspace{5mm}
\leftline{\bf 2.5. Dissipation at $\tau_T < 1$}
\vspace{3mm}

\noindent 
The high energy emission in blazars covers a wide range of energies,
up to 10 GeV (or higher in the TeV sources) and hence
must be powered by non-thermal particle distributions.   The most familiar
possibilities are first-order Fermi acceleration at shocks (Blandford
and Eichler 1987) and electrostatic acceleration (e.g. RL92).
Photo-pion production on protons can more easily 
accomodate the TeV sources (Mannheim 1993), but requires a supplementary
$\sim$ MeV emission mechanism in sources 
with soft high energy spectral states, because of the much greater
cross-section for $\gamma + \gamma \rightarrow e^\pm$.

The relevant physics is treated in sufficient depth elsewhere that
I will focus on two questions here.

1. {\it Does reconnection deposit energy primarily in thermal or
non-thermal particles, and in electrons or ions?\/}  This problem is
far from being understood from first principles, but Solar flares
do provide clear evidence that the efficiency of electron acceleration
can (at least in a non-relativistic plasma) be quite high.  However,
Type III radio bursts (which are powered by flare particles that
escape the Sun along open magnetic field lines) also provide direct evidence
that most of the flare energy is dissipated in the form of {\it bulk
heating\/} of electrons to energies of 10$-$100 keV, with only a small
fraction being deposited in a relativistic, non-thermal tail (Lin 1990).

2. {\it What is the effectiveness of electrostatic and
strong-wave acceleration in a relativistic, Comptonizing medium, 
as compared to shock acceleration?\/}  This question highlights
a crucial difference between the large scale relativistic outflows
associated with blazars and GRBs, and a laboratory system such as a
Tokamak (or even smaller scale magnetic features astrophysical systems such as
coronal loops and arcades).  A~magnetic field $B$ with gradient scale
$\ell_B = B/|{\bf\nabla} B|$ requires a minimal charge density
$n_{c,\rm min} = B/4\pi e\ell_B$ to support the associated current; otherwise
the displacement current cannot be neglected and charges are accelerated
to relativistic energies.  The ratio of the actual electron density
to $n_{c,\rm min}$ can be expressed in terms of the scattering depth across
$\ell_B$,
	\begin{equation}
{n_e\over n_{c,\rm min}} \sim {\tau_T\over\alpha_{em}}\,
\left({B\over B_{QED}}\right)^{-1}.
	\end{equation}
This works out to $n_e/n_{c,\rm min} \sim 10^{13}$ for parameters appropriate to
inhomogeneous external Compton blazar models (BL95).
It has been hypothesized that the boundary layers of jets may
be partially evacuated and sites for strong-wave acceleration 
(Bisnovaty-Kogan \& Lovelace 1997), but in fact the degree of evacuation
must be extraordinary for such a mechanism to be important.  

Another approach to this problem
is to re-express $B$ in terms of the plasma $\beta_e=8\pi n_eT/B^2$,
	\begin{equation}
{n_e\over n_{c,\rm min}} \sim \beta_e\left({B\over B_{QED}}\right)
\left({T\over m_ec^2}\right)^{-1}{\ell_B\over \ell_{m_e}},
	\end{equation}
where $\ell_{m_e} = 4\times 10^{-11}$ cm is the Compton wavelength
of the electron.  This is $n_e/n_{c,\rm min} \sim 10^4\beta_e$,
$10^8\beta_e$, and $10^{21}\beta_e$ for parameters appropriate to 
Tokamaks, Solar flares, and blazar jets.  This suggests that 
much higher wavenumber distortions of the magnetic field are
required to provide efficient electron acceleration through
reconnection in relativistic outflows.
By contrast, direct Comptonization of a background photon
fluid is {\it more\/} effective in jets and GRBs due to the
higher scattering depth (T94).

\vspace{8mm}
\leftline{\large\bf 3. Spectral consequences}
\vspace{5mm}

\noindent 
Most modelling of high energy emission from relativistic outflows 
ignores the inner boundary condition on the outflow.
The flow is hypothesized to dissipate outside the pair
annihilation radius, and flow conditions interior to that radius are
assumed not to influence the emergent high energy spectrum.  Seed
radiation for Comptonization is assumed to originate in a central
accretion disk {\it exterior\/} to the volume of the outflow.

We have already seen, however, that the emergent spectrum can 
be dominated by advected radiation if the outflow is optically
thick at its base (T94, T96).  Moreover, the inner boundary conditions are
very well defined in flows of large central compactness, such
as GRBs ($\ell_c \sim 10^{15}$).   The mean photon energy emerging from 
the flow is $\langle h\nu\rangle \sim L_P/\dot N_\gamma$, when the
asymptotic Lorentz factor lies near the critical value $\gamma_{e^\pm}$
(or $\gamma_{e-p}$ if pairs are absent). Thermalization is rapid near
the base of the flow,  the photon gas is very close to black body, and 
$\langle h\nu\rangle$ is directly related to the effective temperature
at the light cylinder, 
	\begin{equation}
\langle h\nu\rangle \sim T_{\rm eff} = 
0.8\left({L_{\gamma}\over 10^{50}}\right)^{1/4}
\left({P\over{\rm 10^{-3}~s}}\right)^{-1/2}~~~{\rm MeV}.
	\end{equation}
This is remarkably close to the observed range of spectral break
energies, after allowing for cosmological redshift.  

In this regime, the ratio of photon luminosity $L_\gamma$ 
to (ordered) Poynting luminosity $L_P$ at the base of the wind is a key
 parameter.  It should be emphasized that the 
baryon loading is tolerably small only if $L_\gamma < 10^{-2} L_P$ at
the neutrinosphere.   In other words, a key requirement of this model is that 
the wind be reheated from $L_\gamma \ll L_P$ to $L_\gamma \sim L_P$
well outside the neutrinosphere.  This is plausibly accomplished by
a MHD cascade to high wavenumber, even though photons and pairs
are tightly coupled on macroscopic scales (Sect. 2.4; TB97).

Expression (12) also leads to an interesting question:  how does the mean
energy per photon change as one decreases the central compactness?  
As we now show, the trend
of decreasing mean photon energy with decreasing compactness in fact
can be reversed, with $\langle h\nu\rangle$ approaching $\sim m_ec^2$
at $\ell_c \sim 10^2$.

\vspace{5mm}
\leftline{\bf 3.1. Direct MeV Wien peak}
\vspace{3mm}

\noindent 
A magnetized outflow that is strongly turbulent will trigger a
cascade to high wave\-number that {\it must\/} dissipate inside the Alfv\'en
radius.  We look for an optically thick equilibrium state in which the
dissipative zone is shielded from external photons and soft photons are
generated internally.  In applications to AGN the advected matter contributes
negligible optical depth, and so the outflow is necessarily hot and
pair loaded.   Damping via resonant couplings between high wavenumber
turbulence and cosmic ray particles has been considered by Dermer, Miller,
\& Li (1996).  However, direct Compton drag of bulk turbulent motions
is an effective damping mechanism at high compactness, typically at much
lower wavenumbers (TB97).  I now estimate the temperature $T_0$
of the flow at its base, focussing on two photon sources.
\smallskip 

{\it 1. Double Compton Emission.}  This dominates bremsstrahlung emission
when $T_0 \sim m_ec^2$ and the outflow is photon rich, $n_\gamma \gg n_e = 
n_{e^+} + n_{e^-}$.  In a Wien photon gas,
	\begin{equation}
\dot n_{dC} = {16\Lambda\over\pi}\alpha_{em}n_en_\gamma\sigma_T c
\left({T_0\over m_ec^2}\right)^2,
	\end{equation}
where $\Lambda \simeq \ln(T/h\nu_{\rm min})$ and the photon gas approaches
a Planckian distribution at frequency $\nu_{\rm min}$.

Double Compton dominates cyclotron emission when the magnetic
field in the central engine exceeds $B_{QED} = 4.4\times 10^{13}$ G, 
so that thermal pairs do not populate excited Landau levels.
Such strong fields have indeed been associated with SGR 0526$-$66, which emitted
the March 5, 1979 superburst.  The initial 0.1 s pulse of that burst
appears to have approached a luminosity of $\sim 10^7$ times the Eddington
luminosity (Fenimore et al. 1996), and has the appearance of an expanding
pair fireball (Thompson \& Duncan 1995; Fatuzzo \& Melia 1996).  
If the outflow is driven by a~magnetic field that is also strongly
turbulent, then Compton drag can raise the photon energy density 
$3T n_\gamma$ close to $B^2/8\pi$ near the light cylinder (TB97).
Equating $\dot n_{dC}$ with the photon loss rate,
one deduces an equilibrium scattering depth
	\begin{equation}
\tau_T = {\pi\over 16\Lambda\alpha_{em}}\left({T\over m_ec^2}\right)^{-2}.
	\end{equation}
Re-expressing this in terms of the equilibrium pair density yields
the relation between $T_0$ and compactness $\ell_c$ shown in Fig.~4.
\smallskip 

{\it 2.  Cyclo-Synchrotron Emission in a Non-thermal Pair Plasma}.
Such a very high optical depth and compactness cannot be maintained
in weaker magnetic fields.  Cyclo-synchrotron photons are created rapidly,
and their energy rapidly exponentiates.  (For related calculations
with non-thermal particle distributions, see Ghisellini, Guilbert, \&
Svensson 1988.)  To show this, I
parametrize by $N_c eB/m_ec$ the critical frequency at which Compton
scattering increases the frequency of a C-S photon at the same rate as it is
absorbed, $dy/dt = c\alpha_\nu$.  From Kirchoff's law,
	\begin{equation}
{\dot n_{{\rm C-S}}\over \dot n_{dC}} \simeq {2N_c^2\over\Lambda}\,\left({B^2/8\pi
\over n_\gamma m_ec^2}\right).
	\end{equation}
Since $N_c \gg 1$ for $3T \sim m_ec^2$  (Mahadevan, Narayan, \& Yi 1996)
this yields $\dot n_{{\rm C-S}} \gg \dot n_{dC}$.

\begin{figure}[h]
\vspace{78mm}
\centerline{
            \psfig{file=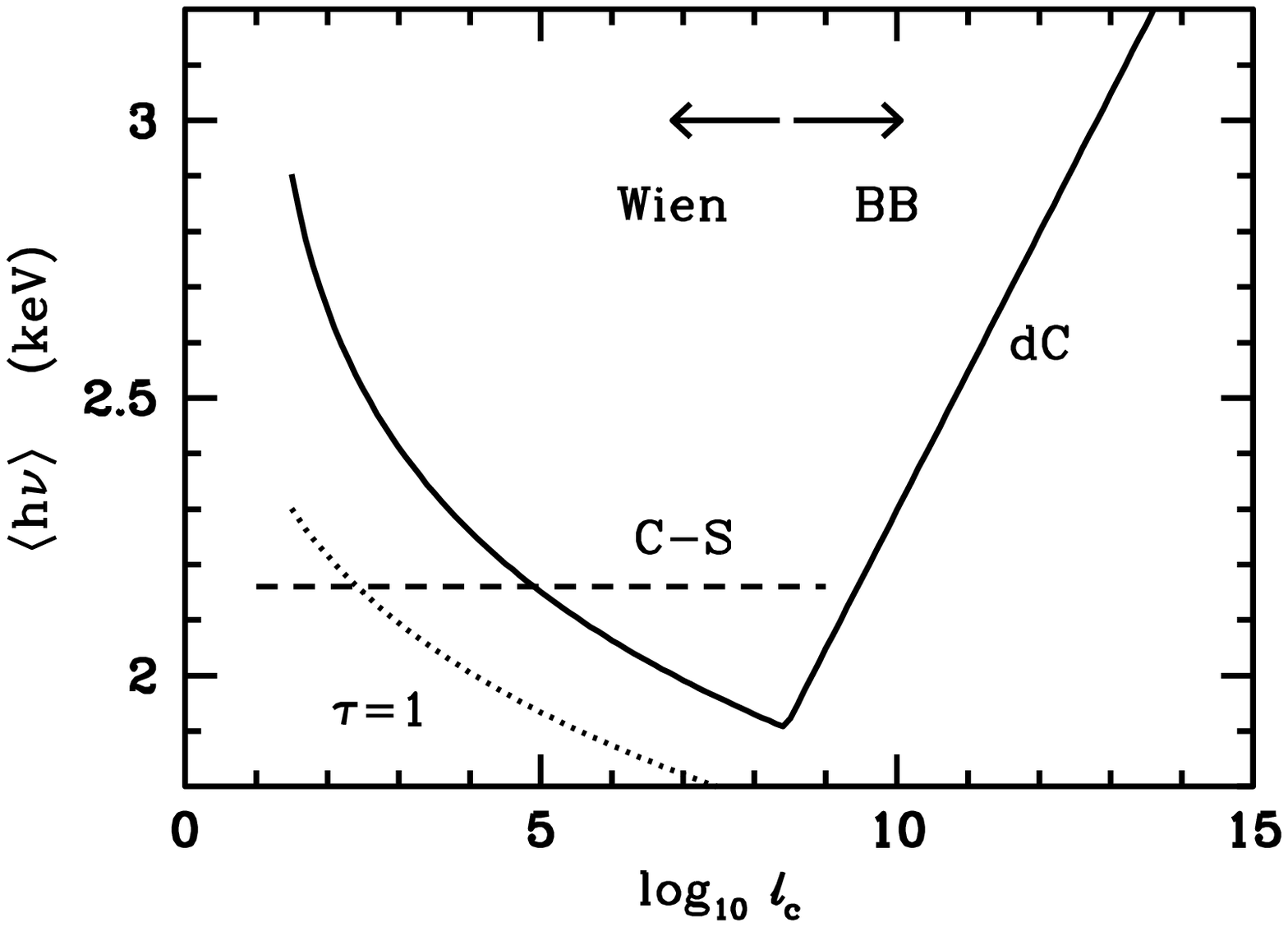,width=10.cm,angle=0}}
\caption{Mean photon energy $\langle h\nu\rangle$ versus central
compactness $\ell_c$
for flows in which double Compton emission (solid line) and
cyclo-synchrotron emission (long-dashed line) is the dominant soft photon
source.  The minimum temperature for an optical thick flow (assuming
a Wien photon distribution) is labelled by the short-dashed line.  Note
the breaks in the curves at the transition from Wien to black-body photon
distributions.}
\end{figure}

Is the energy released by a turbulent cascade
at the base of the outflow dissipated at large or small optical depth?
If even a tiny fraction of the bulk kinetic energy is carried by particles,
then the wave energy must cascade to very high wavenumber before 
charges are accelerated either electrostatically --- as the turbulence
becomes charge starved --- or by resonant interactions.  The relative 
importance of these two effects is determined by the ratio
	\begin{equation}
{k_z\over eB/m_ec^2} \sim \left({\ell\over\ell_{m_e}}\right)^{-1}\,
{\tau_T\over\alpha_{em}}\,\left({B\over B_{QED}}\right)^{-2}
	\end{equation}
in the case of sheared Alfv\'en waves of wavenumber $k_z$ (TB97).
The cascade can, however, be cut off by Compton drag much closer to
the outer scale.  The MHD wave motions are only mildly relativistic,
and so a large scattering depth  is required to provide the Compton parameter
$y \sim 10$ needed to upscatter the bulk of the C-S photons to high energy.
The mean energy per photon is given by
	\begin{equation}
{3\cdot 2T_0\over m_ec^2} \simeq {B^2/8\pi\over m_ec^2 \dot n_{C-S}(R/c)}
\sim {\pi\over 32 N_c^2\alpha_{em}} \tau_T^{-1}\,
\left({T_0\over m_ec^2}\right)^{-2},
	\end{equation}
assuming that thermal and turbulent velocities are comparable and
the bulk of the photon distribution is Wien.   The equivalent temperature
of the (combined) thermal and turbulent motions is 
$2T_0/m_ec^2 \simeq y/4\tau_T$, which gives
	\begin{equation}
\tau_T \sim 20\,\left({y\over 10}\right)^{3/2}\,\left({N_c\over 20}\right).
	\end{equation}
Equivalently, one finds
$2T_0/m_ec^2 = 0.14\,(y/10)^{-1/2}(N_c/20)^{-1}$, as shown in Fig. 4. 

This scattering depth lies above the thermal value at temperature
$2T_0$ unless $\ell_c > \tau_T (n_\gamma/n_e)(6T_0/m_ec^2) \sim 1000$, 
where $n_\gamma/n_e \simeq 300$ in chemical equilibrium.
This represents the
{\it instantaneous\/} compactness of the outflow, and is most easily supplied
by a rapidly rotating Kerr black hole.  However, one can identify
harder photon sources (such as a non-thermal X-ray continuum 
with an energy spectrum peaking at $\sim 100$ keV) that will more easily
seed runaway pair creation as the outflow accelerates
and the turbulent motions become more relativistic.  
On purely energetic grounds the minimum compactness is $\ell_c \sim 10$.



\vspace{5mm}
\leftline{\bf 3.2. Dissipation at large compactness and low $\gamma_\infty$}
\vspace{3mm}

\noindent 
A much wider range of peak energies is possible if enough matter 
is advected by the outflow to make it optically thick.   This is possible
only for a rather higher compactness, $\ell_c > m_p/m_e \sim 2000$,
than is expected in blazar sources, but well within the range of
cosmological GRB sources (cf. Paczy\'nski 1990).  
If the matter loading is heavy enough that $\gamma_\infty 
< \gamma_{e-p} = 0.2 \ell_{\Delta t}^{1/5}$, then dissipation
on an observed timescale $\Delta t$ occurs inside the $e-p$
scattering photosphere.   This provides a natural explanation for the
soft tails and precursors to GRBs seen by Ginga, and the soft sub-pulses
seen by BATSE (T94, T96).

Unlike pair-dominated outflows with negligible matter (Sect. 3.1)
the emergent temperature is very sensitive to the amount of continuous
heating.  (In the case of spherical flows with radial inhomogeneities,
this is equivalent to a broad power spectrum of inhomogeneities.)
I further assume that the flow expands relativistically at its base,
with $\gamma(r) \propto r^\alpha$ with $\alpha > {1\over 2}$.   Then
inhomogeneities in the flow fall out of causal contact,\footnote{In
contrast to the model of MeV blazars discussed in Sects. 2.2 and 3.1,
in which the collimated, lower-$\gamma$ 
flow is assumed to be heated continuously 
by non-radial Kelvin-Helmholtz instabilities.}  and dissipation
on timescale $\Delta t$ is {\it delayed\/} to radius $R_{\rm diss} \simeq 
2\gamma^2c\Delta t$.  The corresponding luminosity $L(R_{\rm diss})$
is then depleted by adiabatic expansion out to the scattering photosphere.
Assuming that photon-number changing processes freeze out before dissipation
takes place, the emergent temperature is a very strong function of matter
 loading,
	\begin{equation}
{T_{\rm obs}\over T_0} = 0.7\,{L(R_{\rm diss})\over L_\gamma(R_0)}
\left({\gamma\over \gamma_{e-p}}\right)^{10/3}\propto 
(\Delta t)^{2/3}.
	\end{equation}
The correction for adiabatic losses (the last factor in this expression)
has a very strong dependence on $\gamma$, but a much weaker dependence
on the (more directly observable) variability timescale $\Delta t$.
This means that the power spectrum of inhomogeneities must cover a very 
wide range of frequencies for continuous heating to overcome the effects of
adiabatic cooling in a {\it radial\/} flow.  The situation is quite
different in collimated flows, where energy can be continuously
extracted from angular (e.g. Kelvin-Helmholtz) instabilities (Sect. 2.2).

The number of advected photons must also be compared with
the number of fresh cyclo-synchrotron photons generated in the
flow outside the central engine.  Following Sect. 3.1, this is
	\begin{equation}
{\dot n_{{\rm C-S}} t\over n_\gamma} \sim {8\alpha_{em} N_{sa}^2\over \pi \gamma}
{d y\over d\ln t}\,\left({T\over m_ec^2}\right).
	\end{equation}
Here all quantities refer to the rest frame of the outflow.
The advected photons suppress $\dot n_{{\rm C-S}}$ at large $\tau_T$
by soaking up energy from the electrons and holding down $y$.  We
conclude that $\dot n_{{\rm C-S}}/n_\gamma$
is typically less than unity for high-$\gamma$ outflows.

\vspace{5mm}
\leftline{\bf 3.3. $e^+-e^-$ amplifier in relativistic outflows}
\vspace{3mm}

\noindent 
The efficiency with which the available energy (in shocks, MHD waves, and
reconnecting magnetic fields) is deposited in high energy photons depends
directly on the fraction $\varepsilon_e$ of the energy deposited in
electrons and pairs.  Pair creation (via photon collisions $\gamma + \gamma
\rightarrow e^+ + e^-$) has traditionally been used to constrain
the emission region in high energy sources 
(e.g. Cavallo \& Rees 1978; Baring \& Harding 1997), but one would like
to emphasize an opposing point of view here:  that the high energy photon flux
from a relativistic outflow can be significantly {\it raised\/} by
pair creation, due to an increase in $\varepsilon_e$.   Energization
of positrons by high-harmonic proton synchrotron maser radiation behind a shock
(Hoshino \& Arons 1991) provides an example of such a leptonic
acceleration mechanism.

Pair creation has, of course, been included for some time in models
of BH accretion disk coronae (e.g. Stern et al. 1996 and
references therein), but in that context the formation of a power law
high energy continuum does not depend essentially on the presence of pairs.
For example, direct Comptonization by MHD motions in the corona will
effectively heat the photons, independently of the 
relative amounts of rest energy in baryons and leptons (T94).  
The situation is quite different in a relativistic outflow that expands
sufficiently rapidly at its base that inhomogeneities fall out of causal
contact.  If these inhomogeneities cover a wide range of spatial frequencies
$k_{\rm min} < k < k_{\rm max}$ (as is needed in GRB models to accomodate the
broad power spectra of the bursts) then pair creation at wave number
$k_{\rm max}$ (radius $\sim 4\pi\gamma^2k_{\rm max}^{-1}$) will increase the
radius of the scattering photosphere by a factor 
	\begin{equation}
{R_{\tau=1}^{e^\pm}\over R_{\tau=1}^{e-p}} \sim
\left({m_p \over m_e}\right)\,
\left({E_{br}\over\gamma m_ec^2}\right)^{2-\beta},
	\end{equation}
given a high energy photon index $\beta$.  This {\it pair amplifier\/}
allows a much wider range of wavenumbers to be dissipated directly by
Compton drag, and hence causes regions of the wind with non-thermal
high energy spectra to be significantly brighter than regions with
thermal spectra (T96). 

At this point I should distinguish between pair amplification 
that is {\it linear\/} (the probability of energization $P_e$ of a newly
created pair is the same as that of a seed electron) and {\it non-linear\/}
(the pairs feed back on $P_e$).  For example, the amplifier operating
at a shock is non-linear if fresh pairs all act as suprathermal 
seeds
for first-order Fermi acceleration; whereas it is linear if the pairs
cool down to the temperature of the background thermal plasma before
interacting resonantly with plasma waves.\footnote{The gyroperiod of a
relativistic electron is orders of magnitude shorter than its cooling
time if the magnetic field contributes an appreciable fraction of the
pressure of the outflow.}   
The corresponding leptonic efficiencies are 
		\begin{equation}
\hspace*{-4pt}
	\begin{array}{l}
\displaystyle \varepsilon_e = 
{P_e(1 + 2n_{e^+}/n_p)\over P_p + P_e(1+n_{e^+}/n_p)}~~~({\rm linear});\\[13pt]
\displaystyle \phantom{\varepsilon_e}
= {P_e + \beta\cdot2n_{e^+}/n_p\over P_p+P_e + 
\beta\cdot 2n_{e^+}/n_p}~~~({\rm non-linear}),
	\end{array}
		\end{equation}
for the simplest case of equal proton and lepton temperatures,
and a non-thermal energy spectrum $dN/dE \propto E^{-2}$.
Here, $P_p$ is the probability of energization of a~proton,
$\beta \simeq m_ec^2/{3\over 2}T$, and 
$\varepsilon_e = P_e/(P_e +P_p)$ in the absence of pairs.

The pair amplifier operating at Comptonizing hotspots (Sect. 3.4)
is non-linear in different sense:  pair creation regulates
the high energy index $\beta$ to the appropriate value to yield
a Thomson depth $\tau_T \sim {1\over 4}(T/m_ec^2)^{-1}$ 
within individual hotspots.  
This second-order Fermi acceleration mechanism therefore yields a power-law
high energy spectrum over a wider range of matter loadings
($\gamma_{e-p} < \gamma_\infty < \gamma_{e^\pm} = 4.5\gamma_{e-p}$) than
does synchrotron cooling of shock-accelerated pairs.

Pair amplification via photon collisions is also {\it non-local}:  
photons upscattered above energy $m_ec^2$ in one hotspot will raise
the pair density in another portion of the flow.  A nice example 
is provided by an expanding shell of matter and photons of radial width
$c\Delta t$.  The bulk Lorentz factor of the photon gas is not
limited by the inertia of the matter when $\gamma > \gamma_{e^\pm}$.
Outside a radius $\sim 2\gamma^2c\Delta t$ 
photons with energies greater than $m_ec^2$ in
the wind frame stream ahead of the forward shock, and sidescatter against
seed electrons.  The pair density grows exponentially (at first)
inside a radius $\sim \ell_c R_0$, until $n_{e^+}/n_p$ reaches unity
and the material ahead of the shock is accelerated to a limiting
Lorentz factor $\sim \ell_\gamma^{1/2}$ (for a hard incident photon
spectrum with $\beta = -2$).  The feedback of pair creation on the
structure of a relativistic shock is an interesting problem that has
not been properly addressed.

One immediate spectral consequence of the pair amplifier is a 
suppression of the minimum leptonic Lorentz factor $\gamma_{\rm min}$,
and hence a suppression of the minimum synchrotron frequency\footnote{The
feedback of pair creation on the formation MeV breaks in high energy
synchro-Compton cascades above accretion disks has been considered
by Done, Ghisellini and Fabian (1990).}
$E_{\rm sync}({\rm min}) \sim \gamma_{\rm min}^2 eB/m_ec$.  
In fact, $\gamma_{\rm min} \rightarrow 1$
as the inertia of the pairs becomes comparable to that of the protons.
To give an example of the potential importance of this effect, consider
a Poynting-flux dominated outflow that approaches its limiting
Lorentz factor $\gamma_\infty$.  The high energy photon index is
taken to be $\beta = -2$ out to a rest frame energy $m_ec^2$.  Near the
scattering photosphere of the wind, the cyclotron energy is
	\begin{equation}
\hbar{eB\over m_ec} = {8\pi ec^2\over \sigma_T (2L_Pc)^{1/2}}
\,\gamma_\infty^3 =  0.04\,\left({\gamma_\infty\over 300}\right)^3\,
\left({L_P\over 10^{51}~{\rm erg~s^{-}}}\right)^{-1/2}\;\;\;\;{\rm eV}.
	\end{equation}
The efficiency of electron acceleration can be increased by pair
creation, but at the cost of suppressing $E_{\rm sync}({\rm min})$ far below
the observed range of break energies in GRB spectra.  {\it As a result, 
the primary emission process must be inverse Compton.}

\vspace{5mm}
\leftline{\bf 3.4. Delayed inhomogeneous Comptonization:}
\leftline{\bf\phantom{3.4. }broken power-law spectra with a
thermal photon source} 
\vspace{3mm}

\noindent 
Let us now consider the photon spectrum that results from delayed
reheating of a~relativistic outflow at large $\ell_c$, 
outside the electron-ion photosphere ($\gamma_\infty \sim \gamma_{e-p}$).
At large $\ell_c$, the photons are adiabatically cooled in between the
central engine and the causal contact radius,
where they are reheated to a luminosity $L_\gamma \sim (\delta B/B)^2L_P$
(when the outflow is Poynting-flux dominated). 
The mean photon energy is restored to a value 
	\begin{equation}
\langle h\nu\rangle \sim 0.7 L_\gamma/L_{\gamma 0},
		\end{equation}
given that photon number is conserved at this radius (Sect. 3.2).

As before, we consider a broad power spectrum of inhomogeneities,
$k_{\rm min} < k < k_{\rm max}$; the corresponding (radial) size of a hotspot
is $\Delta \sim \pi/k$.  Let us suppose that wavenumbers $k_\star < k
< k_{\rm max}$ dissipate inside the electron-ion photosphere, and
$k_{\rm min} < k < k_\star$ outside.  Individual hotspots are assumed to
release their energy when the causal propagation distance
$R/2\gamma_\infty^2$ begins to exceed $\epsilon^{-1}\cdot\Delta$.

Hotspots with $k > k_\star$ dissipate when the
scattering depth of the flow is $\tau_T  = k/k_\star$
(due to seed electrons). The scattering depth across an individual spot
$\tau_T^{\rm spot}$ is smaller by $\varepsilon$.  The resultant spectrum
is Wien when $\tau_T^{\rm spot} \gg 1$.  When $\tau_T^{\rm spot} < 1$
(but the flow itself is still optically thick) cold seed photons
escape the spot before being upscattered, and one may use the
standard loss-probability formalism (Shapiro, Lightman, \& Eardley 1976).
Since the seed photons have adiabatically cooled by a~factor
$\sim (2\gamma_\infty)^{-2/3}(kR_0/2\pi)^{2/3}$, the accumulated $y$-parameter
required to upscatter them is large, and the resulting photon index is
	\begin{equation}
\alpha = {1\over 2} - \sqrt{(9/4) + (4/y)} \simeq -1.
	\end{equation}
This power law distribution [extending up to a mean energy (24)],
with a superimposed Wien peak at energy (24), is the net
result of this first stage of Comptonization.  It compares favorably
with the low energy spectra of GRBs (e.g. Cohen et al. 1996).

As dissipation continues at wavenumber $k \sim k_\star$,
Compton drag regulates the $y$-parameter to a value near unity.
Hotspots with temperature $T_w \sim m_ec^2$ will 
upscatter photons above energy $\langle h\nu\rangle$ in a non-thermal tail
that extends to the pair creation threshold in the wind rest frame.  
This in turn greatly amplifies the number of scattering charges,
since photons greatly outnumber electrons in the outflow (by
a~factor $\sim \gamma_\infty (m_p/m_e) (m_ec^2/\langle h\nu\rangle)$.  

The key point here is that the resulting expansion of the
scattering photosphere feeds back directly on the shape of the high
energy continuum (T96).  If the photon 
compactness $\ell_\gamma = L_\gamma\sigma_T/4\pi\gamma^3 m_ec^3 R \gg 1$
at this radius, then a high energy photon index as hard as $\beta = -2$
generates $\tau_T \gg 1$ within individual hotspots, which
in turn prevents the formation of an extended high energy continuum.
For example, if the heating is triggered by reconnection (T94) then
this requires only that $V_A \sim c$ in the wind rest frame,
so that individual reconnection events induce bulk mass motions at
velocities close to the speed of light.  As the compactness drops,
$\beta$ rises to maintain $\tau_T \sim {1\over 4} y (T_w/m_ec^2)$ 
within individual hotspots.\footnote{
As long as $\epsilon \ll 1$ photons are able to diffuse freely between
spots, but not escape the wind entirely.}  In other words, the feedback
works primarily through the scattering depth, rather than through
a balance between the time-averaged heating and cooling rates ($y = 1$)
as in accretion disk corona models (Shapiro et al. 1976; 
Haardt \& Maraschi 1993).

The net result is that hotspots in the wind with 
the right properties to generate pairs are observed to be much
brighter in X-rays and $\gamma$-rays than are
regions of the wind
with thermal spectra, because a much wider range of wavenumbers
is dissipated by Compton drag (Sect. 3.3).

This mechanism does not require fine-tuning of the Lorentz factor if 
the range of wavenumbers is broad, $k_{\rm max} \gg k_{\rm min}$.  Nonetheless,
one expects that $\gamma_\infty$ is a strong function of time
in any GRB source involving an optically thick neutron torus or
neutron star that emits neutrinos (T94).  The  neutrino luminosity
plausibly passes through the critical value at which the
neutrino driven mass-loss rate is $\dot M = L_P/c^2\gamma_{e-p}$;
indeed the total Poynting luminosity $L_P$ from a centrifugally
supported torus is limited to $L_P \sim 10^{51}$ erg s$^{-1}$ in
this manner.  

A related model uses strong shocks to directly accelerate
the photons via the first-order Fermi process (Blandford and Payne 1981).
If the photons pass successively through several strong shocks
separated by adiabatic cooling, then it can be shown that the number
index converges to a value $-1$ (Melrose \& Pope 1993) up to an energy
$\sim L_P/\dot N_\gamma$.

\vspace{5mm}
\leftline{\bf 3.5. A hybrid model:}
\leftline{\bf\phantom{3.5. }Comptonization of an MeV bump by
non-thermal pairs} 
\vspace{3mm}

\noindent 
An advected Wien photon gas with temperature $T_0 \sim m_ec^2$ can
seed Comptonization by non-thermal pairs below the scattering
photosphere.   If the distribution of relativistic pairs has a lower cutoff
$\gamma_{\rm min} = O(1)$, then {\it the resultant spectrum
breaks in the MeV range}.  Such a low cutoff results from
a high energy pair cascade in a compact photon source (BL95), and 
results even for steeper pair spectra if pair creation feeds back
on the leptonic acceleration efficiency to load the outflow heavily
with pairs, $n_e m_ec^2 \sim (\delta B)^2/8\pi$ (Sect. 3.3).  

A further benefit of heavy pair loading is that the outflow becomes
photon-starved when $3T$ approaches $m_ec^2$ in the rest frame,
so that the advected bump is strongly depleted by Compton upscattering above
$\sim 1$ MeV.   The resultant break energy is then (for $\nu F_\nu \sim $
const above 1 MeV)
	\begin{equation}
h\nu_{br} \sim {L_\gamma\over \dot N_\gamma}\,
\left[\ln\left({h\nu_{\rm max}\over m_ec^2}\right)\right]^{-1}
\sim 3T_0\,\left({L_\gamma\over L_{\gamma 0}}\right)
\left[\ln\left({h\nu_{\rm max}\over m_ec^2}\right)\right]^{-1}.
	\end{equation}
Indeed, a narrow bump near 1 MeV appears to be the exception rather
than the rule in blazar spectra, although PKS 0208$-$512 does provide
a spectacular exception (von Montigny et al., these proceedings).

As a model for blazar spectra, this has a number of advantages over
models involving i) direct Comptonization of photons from the central
source (Dermer and Schliekieser 1993); and ii) Comptonization of
side-scattered photons (Sikora, Begelman, \& Rees 1994; BL95).
First, a Comptonized UV bump (Sikora et al. 1997) is avoided because
the flow is self-shielding
(Sect. 2.2); second, the advected Wien peak has a high enough temperature that
the photon source is depleted during creation of the power-law
$\gamma$-ray spectrum; and, third, observations of {\it both\/} MeV 
power-law breaks and isolated MeV bumps in blazar spectra are
directly tied to the electron rest energy.  {\it The duality between these
two spectral states is ascribed to the presence or absence of a strong
non-thermal $e^\pm$ component.}

Comptonization of an advected MeV bump can be powered either by thermal or
non-thermal particles.  Is it reasonable to expect that quasi-thermal
motions should be the dominant Compton heat source in GRB sources,
but non-thermal particles in blazars?   The key different between these
sources, aside from the central compactness, appears to be 
the degree of relativistic
expansion.  Shocks in a GRB outflow should (locally) more closely approximate
spherical surfaces, with the result that first-order Fermi
acceleration is strongly suppressed in the relativistic limit (cf.
Kirk, these proceedings).

\vspace{4mm}
\leftline{\large\bf 4. Conclusions:  optically thick vs. thin sources}
\vspace{2.5mm}

\noindent 
We have studied dissipation in relativistic outflows that
are sufficiently compact ($\ell_c > 10$) to be optically
thick at the center.  Advected MeV radiation  provides an
interesting new source of Compton seeds in this regime.  Interacting with
bulk turbulent motions and non-thermal pairs {\it near the scattering
photosphere}, it can manifest itself either as an MeV blazar, or as an extended
power law state when most of the available energy is converted to
non-thermal pairs.  The absence of a prominent MeV bump
in most blazar sources can be explained since the outflow is
{\it photon starved}.   The spectral signature is expected to be
different when $\ell_c < 1-10$, or when the advected radiative flux is low.
The high energy spectral break energy can cover a wider range of
frequencies when synchrotron photons are the dominant seeds
(MRP94; Ghisellini, these proceedings; Takahara, these proceedings).

Electron-positron pairs play a crucial role here by
i) maintaining a large scattering depth near the base of the outflow
and shielding the high-$\gamma$ core of a jet from ambient radiation;
ii) maintaining the mean energy of the advected radiation above $\sim 1$ MeV; 
iii) enhancing the efficiency of leptonic dissipative modes;  and iv)
reducing the minimum energy of the non-thermal pair population to 
$\gamma_{\rm min} \sim 1$, which keeps the minimum energy of the Comptonized
MeV photons in the MeV range.  In a large-$\gamma$ (GRB) outflow,
pairs also feed back on the emergent spectrum
by expanding the \mbox{scattering} photosphere, and thus greatly increasing the
range of wavenumbers that are damped by Compton drag off advected
radiation.

{\it Heavy Matter Loading and GRB Afterglow.} 
One should also consider the effects of matter opacity in GRB outflows
with extremely high central compactness.  Although a~core of the outflow
(e.g. near the rotation axis of the central engine) must attain very high
$\gamma_\infty \sim 100{-}300$, material off axis may not.\footnote{For example,
if the central engine is a rapidly-rotating neutron star or neutron torus,
then neutrino emission can easily power mass loss rates as high
as $10^{51}~{\rm erg~s^{-1}}/c^2 \sim 10^{-3}$ $M_\odot$ s$^{-1}$.}
Material expanding with $\gamma_\infty\sim 1{-}2$ becomes optically thin
to scattering on a timescale $\sim 1$ day $(E/10^{52}{~\rm erg})^{1/2}$.
This is comparable to the timescale on which the optical afterglow
detected from GRB970508 reached a maximum (Bond 1997).  This is telling,
since direct synchrotron emission between the contact discontinuity and
the forward shock probably is strongly suppressed due to the relative
weakness of the ambient magnetic field (Sect. 2.2).
A~further motivation for {\it simultaneous\/} optical observations of GRBs
comes from the observation that the minimum frequency $N_c eB/m_ec 
\sim (10^2{-}10^3)\,eB/m_ec$ for optically thin cyclo-synchrotron emission lies
near $\sim 1$ eV (Sect. 3.3) near the $e^\pm$ scattering photosphere
and at a bulk Lorentz factor of $\sim 10^2$.  

{\it GRB Time Profiles: FRED vs. Chaotic.} 
The light-curves of GRBs show a bewildering variety of shapes
(Meegan et al. 1996), but at
least two main classes can be identified:  bursts with smooth, asymmetric
pulses (`Fast Rise and Exponential Decay' or FRED); and more `chaotic' bursts
in which narrow and wide pulses are often superimposed and in which
the asymmetry of individual pulses is usually less clearly defined.  

Chaotic bursts are most easily explained if the dissipation in a burst
is driven by local physics within the outflow (PX94; T94; RM94; Sari \&
Piran 1997), rather than by interaction with an external medium.  This
leads to a simple discriminant between the two classes:  FRED bursts
arise from shells of ejecta that come into causal contact before
the $\gamma$-rays escape, and vice versa for chaotic bursts.  
However, if the FRED bursts are also powered by local physics on
a scale smaller than the width of the shell of ejecta, then the smoothness
of the lightcurves indicates that {\it the $\gamma$-ray emitting zone lies at
scattering depths $\tau_T >  1$}.  Given the lack of a clear spectral
distinction between the two classes, one reaches the same conclusion for
chaotic bursts.  Indeed, the transition zone between
large and small $\tau_T$ can be considerably broadened by pair creation,
and pairs are most effective at enhancing leptonic dissipative modes
near the scattering photosphere (Sect. 3.3). In sum, this leads to the following
simple model:  a~burst is smooth (FRED) or chaotic depending on
whether the {\it scattering photosphere\/} lies outside or inside the causal
contact radius $2\gamma_\infty^2 c\Delta t$.  

{\it Range of Temporal Frequencies and GRB Soft Tails.} 
A flow will, in general, be variable on
a range of timescales $\Delta t$ and so dissipation can occur
over a range of optical depths.  The emergent spectrum varies 
considerably depending on whether most of the available energy
resides at long timescales or short.  

In this regard, it is interesting to note
that the extended soft bump following GRB 870303 (a chaotic burst) detected
by Ginga had a quasi-thermal cutoff (Yoshida et al. 1989), 
whereas the extended soft emission in GRB 960720 (a FRED burst) detected by 
BeppoSAX is closer to an extended powerlaw (with the possibility of a 
cutoff at $\sim 30$ keV; Piro et al. 1997).  This spectral difference could be
explained if the ejecta that produced the soft tail of GRB 870303
dissipated at $\tau_T^{e-p} \gg 1$ while they were causally disconnected
from the primary pulse of ejecta.  High energy cut-offs to GRB afterglow
are in general very diagnostic:  the high energy spectral index is
attracted to $\beta = -2$ below an energy $\sim (\gamma/\tau_T)m_ec^2$ (T94),
and so the presence of a high energy spectral break yields information
about a combination of bulk Lorentz factor and scattering depth (see also
Baring \& Harding 1997).

\vspace{7mm}

\leftline{\large\bf Acknowledgments}
\vspace{3.5mm}

\noindent 
The author would like to thank Michal Ostrowski for comments
on the manuscript, and the Alfred P. Sloan foundation for
financial support.

\vspace{8mm}
 \leftline{\large\bf References}
\vspace{5mm}

\references 

Arav, N.  Begelman, M.C. 1993, ApJ, 413, 700

Baring, M.G.,  Harding, A.K. 1997, ApJ 481, L85

Bisnovaty-Kogan, G.S.,  Lovelace, R.V.E. 1997, preprint

Blandford, R.D.,  Payne, D.G. 1981, MNRAS, 194, 1041

Blandford, R.D.,  Eichler, D. 1987,  Phys. Rep., 154 1

Blandford, R.D.,  Levinson, A. 1995, ApJ 441, 79 (BL95)

Bloemen, H., et al. 1995, A\&A, 293, L1

Blom, J.J., et al. 1995, A\&A, 298, L33

Bond, H.E. 1997, IAU circular 6654; see also circulars 6655-6658

Cavallo, G.,  Rees, M.J. 1978, MNRAS, 183, 359

Cohen, E., Katz, J.I., Piran, T., Sari, R., Preece, R.D., 
Band, D.L. 1996, preprint.

Dermer, C.D.  and Schlickeiser, R. 1993, ApJ, 416, 458

Dermer, C.D., Miller, J.A.,  Li, H. 1996, ApJ 456, 106

Done, C., Ghisellini, G.,  Fabian, A.C. 1990, MNRAS, 245, 1

Duncan, R.C., Shapiro, S.L.,  Wasserman, I. 1986, ApJ 309, 141

Duncan, R.C.,  Thompson, C. 1992, ApJ 392, L9

Fatuzzo, M.,  Melia, F. 1996, ApJ 464, 316

Fenimore, E.E., Klebesadel, R.W.,  Laros, J.G. 1996, 
ApJ, 460, 964

Ghisellini, G., Guilbert, P.W.,  Svensson, R. 1988, ApJ, 334, L5

Haardt, F.,  Maraschi, L. 1993, ApJ 413, 680

Hoshino, M.,  Arons, J. 1991, Phys. Fluids B, 3, 818

Hurley, K., et al. 1994, Nat., 372, 652

Jaroszynski, M. 1993, A\&A, 43, 183

Levinson A. 1996, ApJ, 459, 520

Lin, R.P. 1990, in {\it Basic Plasma Processes on the the Sun},
ed. E.R. Priest \& V. Krishan, p. 467

Mahadevan, R., Narayan, R.,  Yi, I. 1996, ApJ 465, 327

Mallozzi, et al. 1995, ApJ 454, 597

Mannheim, K. 1993, A\&A 269, 67

Maraschi, Ghisellini,  Celotti, 1992, ApJ 397, L5

Marcowith, A., Henri, G., Pelletier, G. 1995, MNRAS, 277, 681

McNaron-Brown, et al. 1995, ApJ 451, 575

Meegan, C.A., et al. 1996, ApJ 106, 65

Melrose, D.B.,  Pope, M.H. 1993, Proc. Ast. Soc. Aust., 10, 222

M\'esz\'aros, P., Rees, M.H.,  Papathanassiou, H. 1994, ApJ 432, 181
(MRP94) 

M\'esz\'aros, P.,  Rees, M.J. 1997, ApJ 482, 29

Paczy\'nski, B. 1990, ApJ, 363, 218

Paczy\'nski, B.,  Xu, G. 1994, ApJ 427, 708 (PX94)

Papathanassiou, H.,  M\'eszar\'os, P. 1996, ApJ 471, L91

Pendleton, G.N., et al., 1996, in proceedings of the
Third Huntsville Symposium on Gamma-Ray Bursts, ed. C. Kouveliotou, M.S. Briggs
\& G.J. Fishman, p.~228

Piran, T.,  Shemi, A. 1993, ApJ 403, L67

Piran, T.,  Narayan, R. 1996, in proceedings of the
Third Huntsville Symposium on Gamma-Ray Bursts, ed. C. Kouveliotou, M.S. Briggs
\& G.J. Fishman, p. 233

Piro, L., et al. 1997, preprint

Rees, M.J.,  M\'esz\'aros, P. 1992, MNRAS, 258, 41

Rees, M.J.,  M\'esz\'aros, P. 1994, ApJ 430, 93 (RM94)

Romanova, M.M., Lovelace, R.V.E. 1992, A\&A, 262, 26

Sari, R., Piran, T. 1997, ApJ, 485, 270

Shapiro, S.L., Lightman, A.P., Eardley, D.M. 1976, ApJ, 204, 187

Sikora, M.,  Begelman, M.C.,  Rees, M.J. 1994, ApJ 421, 153

Sikora, M., Madejski, G., Moderski, R., Poutanen, J. 1997, ApJ
484, 108

Stern, B., Poutanen, J., Svensson, R., Sikora, M., Begelman, M.C.
1995, 449, L13

Thompson, C. 1994, MNRAS, 270, 480 (T94)

Thompson, C., Duncan R.C. 1995, MNRAS, 275, 255

Thompson, C. 1996, in proceedings of the
Third Huntsville Symposium on Gamma-Ray Bursts, ed. C. Kouveliotou, M.S. Briggs
\& G.J. Fishman, p.~802 (T96)

Thompson, C., Blaes, O. 1997, submitted to Phys. Rev. D (TB97)

Usov, V.V. 1992, Nat., 357, 472

Usov, V.V. 1994, MNRAS, 267, 1035

Vietri, M. 1996, ApJ 471, L95

Yoshida, A., et al. 1989, P.A.S.P., 41, 509 

\end{document}